\documentclass[
journal=jctcce, 
manuscript=article]{achemso}
\usepackage[version=3]{mhchem} 
\usepackage{float}
\usepackage{relsize}
\usepackage{color,soul}
\usepackage{amsmath}
\usepackage{flexisym}
\usepackage[table]{xcolor}
\usepackage{booktabs}
\newcolumntype{K}[1]{>{\arraybackslash}p{#1}}

\usepackage{multirow}
\usepackage{booktabs}
\author{Hung Q. Pham}
\author{Matthew R. Hermes}
\author{Laura Gagliardi}
\email{gagliard@umn.edu}
\affiliation[University of Minnesota]
{Department of Chemistry, Chemical Theory Center, and Supercomputing Institute, University of Minnesota, 207 Pleasant Street SE, Minneapolis, Minnesota 55455, United States.}

\title[\texttt{achemso} demonstration]
{Periodic Electronic Structure Calculations With Density Matrix Embedding Theory}

\begin{document}

\begin{abstract}
We developed a periodic version of density matrix embedding theory, DMET, with which it is possible to perform electronic structure calculations on periodic systems, and compute the band structure of solid-state materials. Electron correlation can be captured by means of a local impurity model using various wave function methods, like, for example, full configuration interaction, coupled cluster and multiconfigurational methods. The method is able to describe not only the ground-state energy but also the quasiparticle band picture via the real-momentum space implementation. We investigate the performance of periodic DMET in describing the ground-state energy as well as the electronic band structure for one-dimensional solids. Our results show that DMET is in good agreement with other many-body techniques at a cheaper computational cost. We anticipate that periodic DMET can be a promising first principle method for strongly correlated materials.
\end{abstract}
\section{Introduction \label{sec_intro}}

An accurate and affordable numerical method for strongly correlated electrons in solid-state materials remains one of the most exciting but challenging topics in computational chemistry and material science.\cite{Kent348} This is crucially important because electron correlation governs many exotic phenomena in condensed phases, such as metal-insulator transition, unconventional superconductivity, and magnetism.\cite{strongly_corr_mats1,highT_super,MIT_mott} For decades, Kohn-Sham density functional theory (KS-DFT)\cite{HK_theorem,KS-DFT} has been the most successful method for solid-state materials due to its simplicity and predictive capability for many cases.\cite{DFT_mat1,DFT_mat2} While formally exact, the practical application of KS-DFT using approximate exchange-correlation (XC) functional is unable to provide a good description for strong electronic interaction. This is often attributed to the single-determinant nature of the KS fictitious system.\cite{DFT_challenges} Even within the weak correlation regime, the \textit{exact} KS orbitals energy gap or simply KS band gap ($\equiv \epsilon_{LUMO} - \epsilon_{HOMO}$, with $\epsilon_{LUMO}$ and $\epsilon_{HOMO}$ are the lowest unoccupied and highest occupied molecular orbital energy, respectively) cannot be interpreted as the fundamental band gap ($\equiv IP - EA$ with $IP$ and $EA$ are ionization potential and electron affinity, respectively) owing to the derivative discontinuity of the exchange-correlation energy ($IP - EA = \epsilon_{LUMO} - \epsilon_{HOMO} + \Delta $, with $\Delta$ is the derivative discontinuity).\cite{Discontinuities_Perdew} Strictly speaking, KS-DFT band structure is unphysical as pointed out in the early work by Perdew and Levy,\cite{Discontinuities_Perdew} Sham and Schluter,\cite{Discontinuities_Sham} and recently by Baerends.\cite{KSgap_Baerends} This argument also applies to approximate functionals based on the local density approximation (LDA) or generalized gradient approximation (GGA). LDA/GGA usually underestimate the fundamental band gap of solids. (For solids the fundamental band gap is very close to the optical band gap which is related to the first excitation;\cite{KSgap_Baerends} the distinction is not necessary here and we use the term 'band gap' to refer to the fundamental band gap throughout this paper.) This so-called band gap problem\cite{bandgap_prob}  makes KS-DFT  less appealing as an accurate band structure method for materials. Hartree-Fock (HF) exchange in hybrid functionals\cite{B3LYP91_bandgap,hybrid_bandgap} or the on-site interaction (U) in DFT+U\cite{LDAU} may improve the performance in certain situations. However, the tuning parameter (HF exchange percentage or U correction) requires a justification by experimental measurements, which are often not accessible for new materials, thereby reducing their applicability to materials design.

Moving beyond DFT, GW\cite{GW_OEP,GW_review} has been widely considered as the method of choice for band gap/band structure predictions for weakly correlated systems. Meanwhile, dynamical mean-field theory (DMFT)\cite{DMFT_lattice} combined with KS-DFT, denoted as DFT+DMFT, remains one of the most accurate methods for strongly correlated materials, such as Mott insulators and correlated metals.\cite{stronglyMat_DMFT,DFT-DMFT-Turan} Both are Green's function-based theories and are able to provide a comprehensive understanding of electronic band structure beyond the mean-field approximation in KS-DFT. However, the frequency-dependent formulation is computationally  very expensive, and these methods are thus affordable only for crystals with small unit cells sampled by a \textbf{k}-mesh with moderate size (\textbf{k} is the crystal momentum). Recently, density matrix embedding theory (DMET) has been proposed as a computationally cheaper alternative to DMFT while offering the same accuracy for lattice models.\cite{DMET, DMET_2DHubbard, DCA-DMET, DMET_Hubbard_Holstein} DMET has been extended to treat ground-state and excited states for several chemical systems using different quantum chemical solvers.\cite{DMET_mol,DMETpractical,DCA-DMET,CASDMET,DMET_excited}. Remarkably, DMET has inspired the development of other theories based on density matrix embedding, such as density embedding theory (DET),\cite{DET} bootstrap embedding,\cite{BE1D,BE2D, BEmol} incremental embedding,\cite{IE} localized active space self-consistent field (LASSCF),\cite{LASSCF,LASSCF_spin} and projected density matrix embedding theory.\cite{p-DMET} Its extensions to the real-time formulation\cite{DMET_realtime} as well as to the electron-phonon interacting system \cite{DMET_phonon, DMET_Hubbard_Holstein} have also been investigated.  For periodic systems, DET with a coupled cluster solver has been implemented to compute correlation energy in several one-dimensional systems.\cite{DET_solid} DET can be seen as a one-shot calculation in the DMET framework in which the mean-field quantum bath is not updated self-consistently. It has been shown in our previous work that the optimization of the bath can significantly improve the accuracy for system with translational symmetry like the hydrogen chain.\cite{CASDMET} To the best of our knowledge, the development of DMET as a band structure method for realistic chemical models of condensed-phase materials, has not yet been explored.

In this work, we extend the DMET method to solid-state materials by making use of the local nature of Wannier functions and the translational symmetry of crystals. Our periodic DMET exploits the dual representation of periodic systems in which a local embedding model is constructed in real space and the mean-field wave function is updated in momentum space in an iterative manner. The proposed method is not only able to study ground-state energy but also non-local properties such as quasiparticle band structures. We investigate the performance of the periodic DMET on a series of one-dimensional solids at different lattice constants, including hydrogen (1D-H), lithium hydride (1D-LiH), and polyyne. Our results show that the ground-state energy by DMET is in an excellent agreement with that of the equivalent non-embedding calculations while working on a much smaller Hilbert space. More importantly, the DMET band structure albeit its simplicity agrees well with those obtained by more complicated many-body techniques. DMET can be seen as a wave function-based alternative to DFT+DMFT, however, being free from double-counting because of the use of HF as the mean-field level, it is an attractive \textit{ab initio} method for strongly correlated materials.

\section{Theory \label{sec_theory}}
We first  discuss the real space-momentum space dual representation for a crystal in Section \ref{sec_theory_dualrep}. We then present key components of the DMET algorithm while highlighting the key differences between molecular DMET and periodic DMET in Section \ref{sec_theory_pdmet}. Section \ref{sec_theory_band} discusses our simple scheme to construct a quasiparticle band structure from DMET. A detailed implementation together with its advantages as well as shortcomings are presented in Section \ref{sec_theory_imp}.

\subsection{Dual representation for periodic systems \label{sec_theory_dualrep}}
In mean-field theories, \textit{e.g.}, HF or KS-DFT, for periodic systems, a ground-state wave function of a perfect crystal is conveniently determined by a set of one-electron crystalline orbitals known as Bloch wave functions, $\psi_{m}^{\mathbf{k}}(\mathbf{r})$, or simply Bloch functions, in the momentum space (\textit{i.e.}, \textbf{k}-space). Since the lattice-translation operator commutes with the Hamiltonian, the crystal momentum \textbf{k} together with the band index \textit{m} ($\textit{m} = 1,2,...,N_{band}$) are good quantum numbers to label the periodic wave function. The relation between the eigenvalues $E_m(\mathbf{k})$ of Bloch functions  and the momentum \textbf{k} is referred to as a dispersion relation or an electronic band structure. It is important to emphasize that the band structure is a direct consequence of the translational symmetry. When this symmetry is disregarded (\textit{e.g.}, in a supercell calculation using $\Gamma$-point sampling\cite{BZsampling}), the band structure is reduced to an orbital energy diagram of molecular systems. Owing to the periodicity, the band structure in the first Brillouin zone (FBZ), \textit{i.e.}, the Wigner-Seitz cell\cite{Wigner-Seitz} of the reciprocal lattice, carries complete information of the wave function of the solid and all the observables can be obtained by integrating over the FBZ using a finite number ($N_k$) of quasimomentum vectors or '\textbf{k}-points'.
Alternatively, one can represent a wave function of a crystal in the real space (\textbf{R}-space) by a set of Wannier functions, $\omega_{n}^{\mathbf{R}}(\mathbf{r})$, where \textbf{R} and \textit{n} are the cell and band index, respectively. Here, the lattice vector \textbf{R}  is used as a cell index to label the wave function. The real and momentum space representations are related by the discrete Fourier transform (DFT)

\begin{equation}
\omega_{n}^{\mathbf{R}}(\mathbf{r}) = \frac{1}{N_k} \sum_{\mathbf{k} \subset 
\mathrm{FBZ}} \bigg[ \sum_m^{N_k} U_{mn}^{\mathbf{k}} \psi_{m}^{\mathbf{k}}(\mathbf{r}) \bigg]  e^{-i\mathbf{kR}}
\label{WF:1}
\end{equation}
\begin{equation}
= \frac{1}{N_k} \sum_{\mathbf{k} \subset 
\mathrm{FBZ}}  \widetilde{\psi}_{n}^{\mathbf{k}}(\mathbf{r}) e^{-i\mathbf{kR}} 
\label{WF:2}
\end{equation}
\begin{equation}
= \mathrm{DFT}  \Big[ \widetilde{\psi}_{n}^{\mathbf{k}}(\mathbf{r}) \Big]
\label{WF:3}
\end{equation}
where $\widetilde{\psi}_{n}^{\mathbf{k}}(\mathbf{r})$ is defined implicitly. The unitary transformation matrix $U_{mn}^{\mathbf{k}}$ are chosen to localize Wannier functions in real space by minimizing a spread functional\cite{MV1997} 
\begin{equation}
\Omega = \sum_n \langle \omega_n^{\mathbf{R_0}}|\hat{\mathbf{r}}^2-\bar{\mathbf{r}}_n^2|\omega_n^{\mathbf{R_0}} \rangle
\label{WF:4}
\end{equation}
where $\bar{\mathbf{r}}_n$ is the WF centroid.

\begin{figure}[H]
    \centering
    \scalebox{.1}{\includegraphics{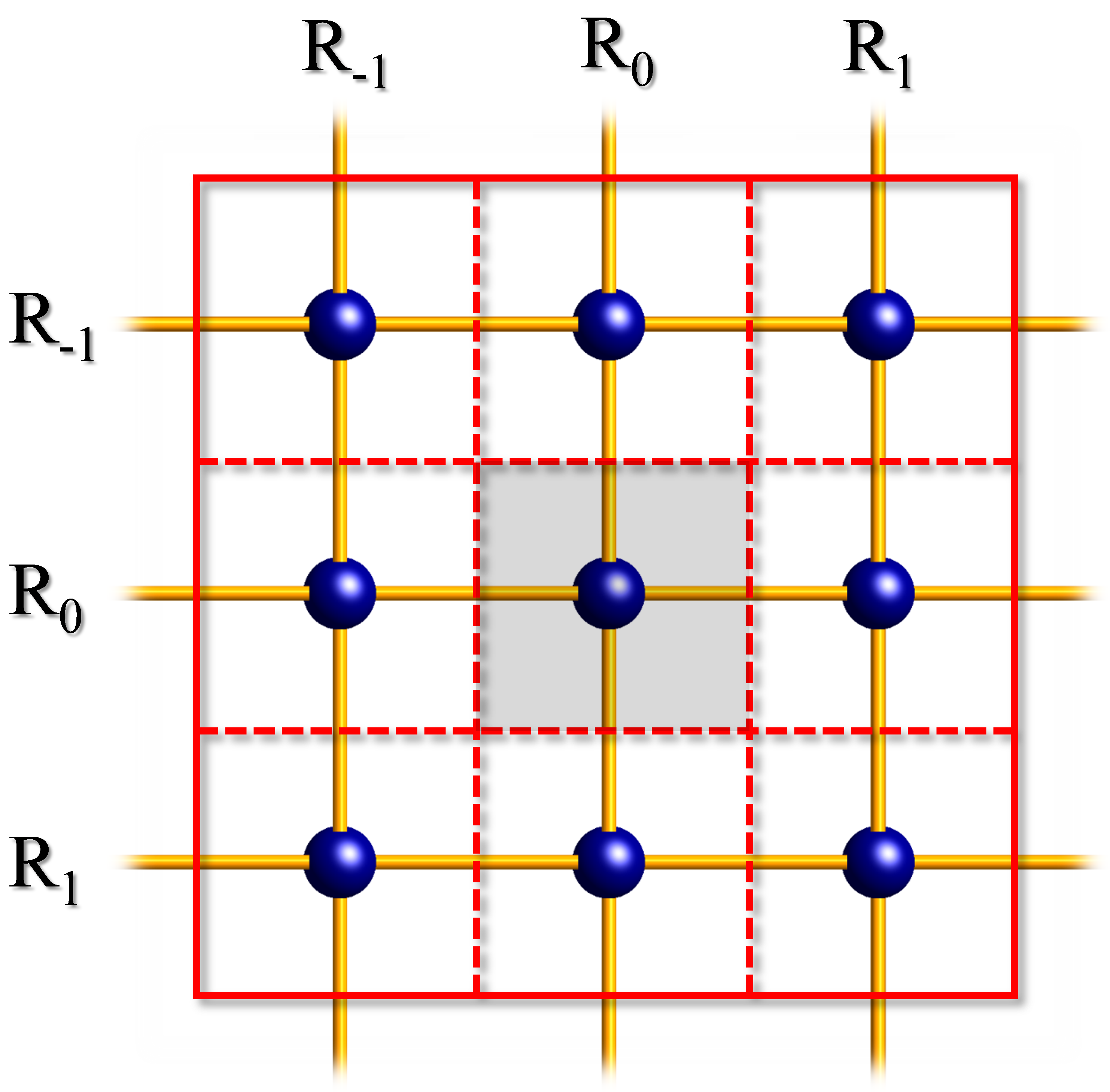}}
    \caption{A 3-by-3 computational supercell subjected to the perodic boundary condition for a square crystal. Each blue ball represents a WF or a group of WFs. $\mathrm{R}_n$ ($n$ = -1, 0, 1) is the cell coordinate and the unit cell colored in grey at the origin is referred as the reference unit cell.}
    \label{supercell}
\end{figure}

The WFs are conceptually similar to the atomic orbitals or basis functions in quantum chemistry. The $U_{mn}^{\mathbf{k}}$ can be computed during a self-consistent field (SCF) calculation\cite{SCF_WFs} or via a post-SCF approach as proposed by Marzari and Vanderbilt\cite{MV1997} [resulting in so-called maximally-localised Wannier functions (MLWFs)]. In this work, we  use the latter procedure to compute $U_{mn}^{\mathbf{k}}$ because this method allows one to straightforwardly construct WFs from any band structure method without the need to modify the underlying SCF algorithm. One can think of a crystal in the real space as a giant molecule subjected to the Born–von Karman boundary conditions\cite{Martin}, defined as a computational supercell, composed of as many unit cells as the numbers of \textbf{k}-points used to sample the FBZ.  Figure \ref{supercell} shows such a computational supercell for a square crystal in a two-dimensional space. 

Generally, one needs to include only a few 'important' bands ($N_{\mathcal{A}} \leq N_{band}$) around the Fermi level in the construction of MLWFs and can leave the rest as frozen bands. The bands around the Fermi level form an 'active' subspace $\mathcal{A}$, which is similar to the active space concept in quantum chemistry. The Hamiltonian in the active subspace, $\hat{H}_{\mathcal{A}}$,
has fewer degrees of freedom than the original Hamiltonian, $\hat{H}$, due to the exclusion of chemically-irrelevant bands. In particular, in real space the one-body part of $\hat{H}_{\mathcal{A}} $ has the size of $(N_{\mathcal{A}} \times N_{k})^2$ compared to $(N_{band} \times N_{k})^2$ of the original problem. Similarly, the two-body part of $\hat{H}_{\mathcal{A}} $ and $\hat{H} $ have the size $(N_{\mathcal{A}} \times N_{k})^4$ and $(N_{band} \times N_{k})^4$, respectively. However, diagonalizing $\hat{H}_{\mathcal{A}}$ by an exact method like full configuration interaction, FCI, for real materials is still practically intractable because one generally needs a large number of $\textbf{k}$-points, which is equivalent to a large computational supercell, to reach the thermodynamic limit (TDL).\cite{TDL} Hence, imposing adequate approximations is crucially important in electronic structure theories for periodic systems. In the next section, we will discuss how the localized nature of MLWFs and translational symmetry can be exploited to construct an impurity model with an even smaller Hamiltonian than $\hat{H}_{\mathcal{A}} $, which can be handled by highly accurate methods.

\subsection{Periodic DMET algorithm \label{sec_theory_pdmet}}
The derivation of DMET has been presented repeatedly in the literature;\cite{DMET,DMET_mol} here, we briefly review its features as related to our periodic implementation. For any system, one can always partition it into two subsystems F, \textit{i.e.}, fragment, and E, \textit{i.e.}, environment with the assumption that the Hilbert space of F is smaller than that of E. F can be a group of sites in a lattice model or a group of atoms in a molecule or a unit cell in a crystal. DMET generates an impurity model containing F and the part of E which is entangled to F by performing a Schmidt decomposition on an approximate single-determinantal wave function, 
\begin{equation}
|\Phi_{\mathrm{tr}}\rangle = \Big( \sum_i^{N_f} \lambda_i |f_i \rangle \otimes |b_i \rangle\Big) \otimes |\mathrm{core} \rangle 
\label{Schmidt_HF}
\end{equation}
where $\lambda_i$ is a coefficient, $|f_i \rangle$ and $|b_i \rangle$ are determinants in the Fock space of fragment and bath orbitals, respectively, and $N_f$ is the number of fragment states. The Schmidt decomposition separates E into the bath ($|b_i \rangle$), which is entangled to the fragment, and the core ($|\mathrm{core}\rangle$), which is unentangled from the fragment, and by construction there can be no more bath states than fragment states. Projecting the Hamiltonian into the impurity basis of $|f_i \rangle$ and $ |b_i \rangle$ generates the impurity Hamiltonian ($\hat{H}_{\mathrm{imp}}$), which has a smaller size than the original Hamiltonian owning to the exclusion of the unentangled core state. This impurity model is similar to an Anderson impurity model \cite{Andersonimpurity} where an infinite lattice problem is mapped to a finite local problem.\\

When the system of interest is a crystal with translational symmetry under periodic boundary conditions, one or more unit cells can be chosen as the fragment and the other unit cells play the role of the environment. However, using two or more unit cells, \textit{i.e.}, an impurity cluster, as a fragment can potentially break the translational invariance within the impurity cluster and a special treatment like the dynamical cluster approximation (DCA) formulation must be used to preserve the symmetry.\cite{DCA-DMET} In our work, we simply restrict the fragment to the reference unit cell in order to preserve the translational symmetry within the computational supercell. We would like to emphasize that the impurity Hamiltonian $\hat{H}_{\mathrm{imp}}$ has a significant smaller size compared to that of  $\hat{H}_{\mathcal{A}} $. Indeed, the one-body part of $\hat{H}_{\mathrm{imp}}$ and $\hat{H}_{\mathcal{A}} $ have the size of $(2 \times N_{\mathcal{A}} )^2$ and $(N_{\mathcal{A}} \times N_{k})^2$, respectively. Similarly, $(2 \times N_{\mathcal{A}} )^4$ and $(N_{\mathcal{A}} \times N_{k})^4$ are the size for the two-body part of $\hat{H}_{\mathrm{imp}}$ and $\hat{H}_{\mathcal{A}} $, respectively. One can easily see that the size of $\hat{H}_{\mathrm{imp}}$ does not depend on the number of \textbf{k}-points used to sample the FBZ. This is the most appealing consequence of DMET algorithm applying to systems with translational invariance like crystals.

In DMET, one needs a set of orthonormal orbitals to define the fragment and the environment. For molecules, these are localized orbitals obtained by some localization schemes, \textit{e.g.}, Foster-Boys,\cite{Foster_Boys} meta-Löwdin, intrinsic atomic orbitals (IAO).\cite{IAO} For periodic systems, it is natural to utilize Wannier functions (WFs) since they are well-localized orbitals satisfying the orthonormality condition
\begin{equation}
\langle  \omega_{m}^{\mathbf{R}} | \omega_{n}^{\mathbf{R'}} \rangle = \delta_{RR'}\delta_{mn}
\label{WF_orthogonality}
\end{equation}
Similar to molecular DMET, the bath orbitals can be constructed by the SVD of the fragment block of the occupied orbital coefficient tensor. This is equivalent to the eigenvalue decomposition of the environment block ($D_{E}$) of the one-body reduced density matrix (1-RDM) of the computational supercell

\begin{equation}
D_{E}  =  \textbf{U} \lambda \textbf{U}^{*}
\label{SVD}
\end{equation}

where  $\lambda$ is the diagonal matrix with $N_f$ none-zero eigenvalues $\lambda_i$. Once the embedding basis, thus the impurity Hamiltonian, is defined, solving the impurity model by a high level-method  is similar to the molecular DMET algorithm. In particular, one needs to solve
\begin{equation}
(\hat{H}_{\mathrm{imp}} + \mu \hat{N} ) |\Psi_{\mathrm{imp}}\rangle  = E_{\mathrm{imp}} |\Psi_{\mathrm{imp}}\rangle 
\label{imp_Hamil}
\end{equation}
where $\hat{N} = \sum \alpha_{p}^{\dagger}\alpha_{q} $ is the particle number operator and $\mu$ is determined so that
\begin{equation}
\langle\Psi_{\mathrm{imp}}|\hat{N} |\Psi_{\mathrm{imp}}\rangle  = N_{\mathrm{cell}},
\label{no_elec_opt}
\end{equation}
where $N_{\mathrm{cell}}$ is the number of electrons per unit cell. Moreover, $\mu = 0$ if one uses the same quantum chemical solver, \textit{i.e.}, HF, to construct the fragment-bath basis as well as to solve the impurity Hamiltonian, \textit{i.e.}, $|\Phi_{\mathrm{tr}} \rangle \equiv |\Psi_{\mathrm{imp}}\rangle$. We use this ``exact embedding''\cite{mol_DMET} to test the fidelity of our Wannierization algorithm.

As standard in DMET, the bath orbitals can be improved by applying a correlation potential $\hat{u} = \sum_{pq} u_{pq}\alpha_{p}^{\dagger}\alpha_{q} $, to the mean-field (low-level) Hamiltonian used to obtain $|\Phi_{\mathrm{tr}} \rangle$ in order to minimize the difference between the 1-RDM of $|\Phi_{\mathrm{tr}}\rangle$ and that of $|\Psi_{\mathrm{imp}}\rangle$. In $\mathbf{k}$-space, we find $|\Phi_{\mathrm{tr}}(\mathbf{k}) \rangle$ by solving the eigenvalue equation
\begin{equation}
\hat{h}({\mathbf{k}}) |\Phi_{\mathrm{tr}} ({\mathbf{k}})\rangle  = \epsilon({\mathbf{k}}) |\Phi_{\mathrm{tr}} ({\mathbf{k}})\rangle 
\label{mf_Hamil}
\end{equation}

\begin{equation}
\hat{h}({\mathbf{k}}) = \hat{F}_{\mathcal{A}}({\mathbf{k}}) + \hat{u} 
\label{fock_k}
\end{equation}

where $\hat{h} ({\mathbf{k}})$ is the low-level Hamiltonian and $\hat{F}_{\mathcal{A}}({\mathbf{k}}) $ is the Fock operator (or the one-electron part of $\hat{H}_{\mathcal{A}} $) projected into the active subspace $\mathcal{A}$ discussed in section \ref{sec_theory_dualrep}. Here, we assume that $\hat{u}$ is not \textbf{k}-dependent; this is equivalent to a block-diagonal form in the real space representation of $\hat{u}$. We choose $\hat{u}$ to optimize the cost function
\begin{equation}
\hat{u} \leftarrow \mathbf{min}_{u_{pq}} \mathbf{CF}(u_{pq}) \equiv \mathbf{min}_{u_{pq}} \Big( D_{\mathrm{emb}}^{l}[u_{pq}] -D_{\mathrm{emb}}^{h}\Big)^2
\label{CF:1}
\end{equation}
where $D_{\mathrm{emb}}^{l}$ and $D_{\mathrm{emb}}^{h}$ are the \textbf{R}-space 1-RDMs associated with the low-level and high-level wave functions in the embedding basis, respectively.
The use of the \textbf{k}-space low-level Hamiltonian to obtain the electronic band structure from the local impurity model will be discussed in section \ref{sec_theory_band}.

\subsection{Electronic band structure from DMET \label{sec_theory_band}}
As discussed in section \ref{sec_theory_dualrep}, an electronic band structure is often modeled as the eigenvalues of the \textbf{k}-space one-electron Hamiltonian, \textit{i.e.}, the Fock operator in the Hartree-Fock theory or the Kohn-Sham operator in KS-DFT. In DMET, following the strategy employed by Reinhard \textit{et al.} in computing fundamental band gaps from DMET calculations on the Hubbard-Holstein model,\cite{DMET_Hubbard_Holstein} we obtain the band structure by diagonalizing a low-level Hamiltonian containing a correlation potential which is optimized by comparing the correlated and mean-field density matrices. However, we have found that the operator $\hat{h}({\mathbf{k}})$ from eq (\ref{fock_k}) utilizing the cost function of eq (\ref{CF:1}) results in unphysically dispersive bands because a change in the cost function in eq \ref{CF:1} can generate a drastic variation of $\hat{h}({\mathbf{k}})$ at certain \textbf{k}-points. We note that the band structure is in fact a global observable corresponding to the total system. In view of this, we use a second, separate effective one-body Hamiltonian to compute the band structure,
\begin{equation}
\hat{h}^{'}({\mathbf{k}}) = \hat{F}_{\mathcal{A}}({\mathbf{k}}) + \hat{u}^{'} 
\label{fock_k_prime}
\end{equation}
where $\hat{u}^{'}$ solves
\begin{equation}
\hat{u}^{'} \leftarrow \mathbf{min}_{u_{pq}} \mathbf{CF}(u_{pq}^{'}) \equiv \mathbf{min}_{u_{pq}^{'}} ( D^{l}[u_{pq}^{'}] -D_{\mathrm{global}}^{h}\Big)^2
\label{CF:2}
\end{equation}
where $D^{l}$ and  $D_{\mathrm{global}}^{h}$ are the low-level and high-level 1-RDM associated with the computational supercell, respectively. Note that $D_{\mathrm{global}}$ depends indirectly on the original correlation potential, $\hat{u}$, which solves eq (\ref{CF:1}). We compute the global correlated 1-RDM as in Ref.\ \citenum{p-DMET}, but applying to crystal systems and utilizing translational symmetry: the rows associated with the reference unit cell ($\mathbf{R}_0$) are obtained from the embedding 1-RDM ($D_{\mathrm{emb}}^{h}$) by transforming back to the WF basis.
\begin{equation}
D_{\mathbf{R_0}}^{h} = (C_{\mathrm{\mathrm{emb}}}D_{\mathrm{emb}}^{h}C_{\mathrm{emb}}^T)_{\mathbf{R_0}}
\label{global_1RDM_R0}
\end{equation}
where $D_{\mathbf{R_0}}^{h}$ is the global correlated 1-RDM  associated with the reference unit cell and $C_{\mathrm{emb}}$ are the coefficient vectors of the embedding orbitals. Next, the rows of the correlated 1-RDM matrix corresponding to the other unit cells are obtained by means of the translational operators
\begin{equation}
D_{\mathbf{R}}^{h} = \hat{T}_R D_{\mathbf{R_0}}^{h}
\label{global_1RDM_R}
\end{equation}
\begin{figure}[H]
    \centering
    \scalebox{1.0}{\includegraphics{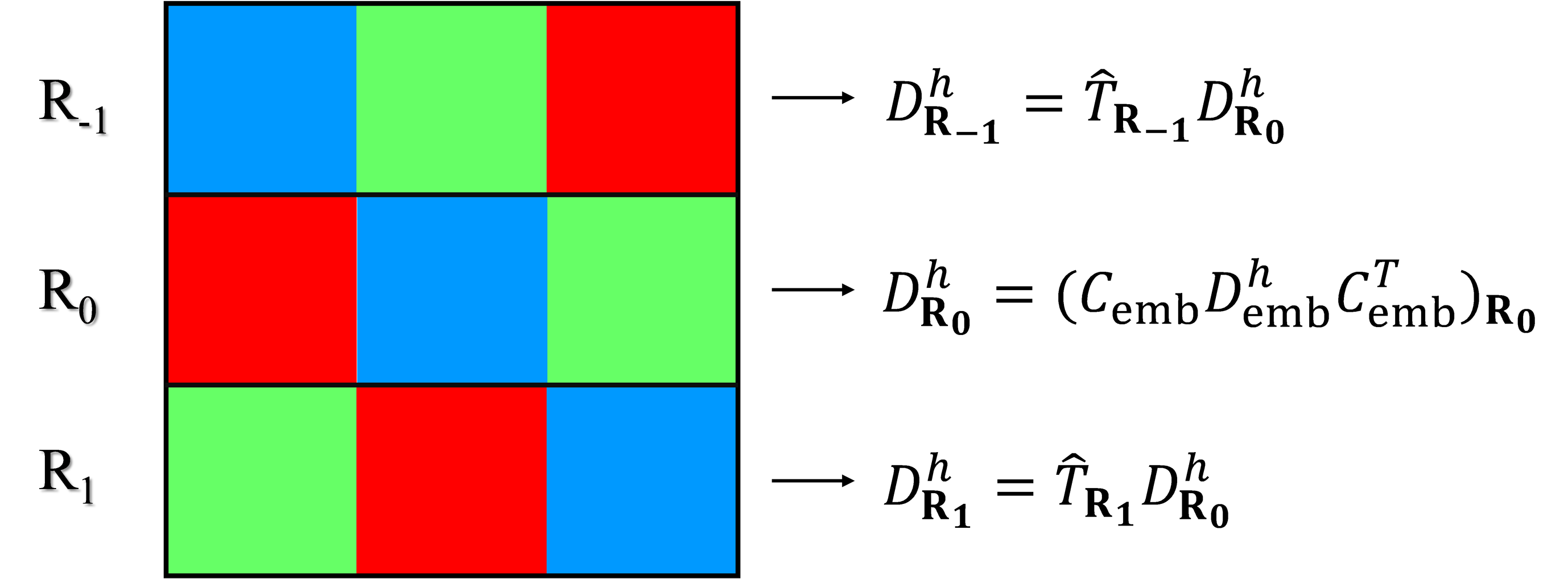}}
    \caption{Construction of a global correlated 1-RDM from the embedding 1-RDM for a 3-by-3 supercell. The 1-RDM for the reference cell are computed from the embedding 1-RDM while those for the other unit cells are obtained by exploiting translational symmetry. $\mathrm{R}_n$ ($n$ = -1, 0, 1) is the cell coordinate.}
    \label{global_1RDM}
\end{figure}

This procedure is summarized by a schematic representation in Figure \ref{global_1RDM}. Finally, we average the resulting 1-RDM with its Hermitian conjugate,
\begin{equation}
D_{\mathrm{global}}^{h} \leftarrow \frac{D_{\mathrm{global}}^{h} + (D_{\mathrm{global}}^{h})^\dagger}{2}
\label{global_1RDM_averaging}
\end{equation}

Once $D_{\mathrm{global}}^{h}$  is computed, we solve eq (\ref{CF:2}) and diagonalize $\hat{h}^{'}(\mathbf{k})$ from eq (\ref{fock_k_prime}) to obtain the band structure. This correlated band structure is an approximate quasiparticle picture. Per Green's function theory,\cite{MarchYoungSampanthar} as invoked in, for instance, DFT+DMFT,\cite{DFT-DMFT-Turan} exact ionization potentials and electron affinities solve the self-consistent eigenvalue equation,
\begin{equation}
    \left(\hat{F}_{\mathcal{A}}({\mathbf{k}}) + \hat{\Sigma}[\mathbf{k},E_m(\mathbf{k})]\right)|\psi_m^{\mathbf{k}}\rangle = E_m(\mathbf{k})|\psi_m^{\mathbf{k}}\rangle
\end{equation}
known as the Dyson equation, where the energy-dependent operator $\hat{\Sigma}$ is known as the self-energy and where the Bloch eigenstates $\psi_m^{\mathbf{k}}$ are the so-called Dyson orbitals.\cite{Goscinski1970,Pickup1973,Cederbaum1975,Ortiz1996} Green's function theory is an \emph{exact} theory - i.e., given the exact self-energy, obtained by summing over an infinite diagrammatic expansion,\cite{MarchYoungSampanthar} this single-particle eigenvalue equation gives exact energy differences between various $N$-electron and $N\pm 1$-electron states. Any approximation to $\hat{\Sigma}$ which omits the energy dependence results in a single unitary transformation of molecular orbitals to Dyson orbitals, implicitly generating a single determinant whose orbital energies yield the approximate band structure. We conjecture that the correlation potential $\hat{u}^{'}$ is an effective energy-independent approximation to the self-energy, and the results of our calculations reported below provide circumstantial evidence to that effect.

\subsection{Periodic DMET implementation \label{sec_theory_imp}}
In this section, we list all the algorithmic steps in a periodic DMET calculation as well as discuss advantages and technical challenges in our theory and current implementation. The DMET algorithm can be summarized as follows:

\begin{enumerate}
\item Perform a HF calculation using a certain \textbf{k}-point mesh, for example, a uniform $\Gamma$-centered or a Monkhorst-Pack mesh.\cite{MPmesh}
\item Define an active subspace $\mathcal{A}$ which includes important bands.
\item Transform the \textbf{k}-space one-electron and two-electron repulsion integrals (1-ERI and 2-ERI), to the \textbf{R}-space.
\item Initialize $\hat{u}$: $u_{pq} = 0$.
\item Get the trial wave function $|\Phi_{\mathrm{tr}}(\mathbf{k}) \rangle$ using eqs \ref{mf_Hamil}-\ref{CF:1}. 
\item Construct the bath orbitals using eq \ref{SVD}.
\item Transform  $\hat{H}_{\mathcal{A}}$ to $\hat{H}_{\mathrm{imp}}$ using the projection operator $\hat{P}= \sum_{i,j}|f_i \rangle \otimes |b_i \rangle \langle f_j|  \otimes \langle b_j  |$.
\item Initialize $\mu$ ($\mu = 0$) and solve $\hat{H}_{\mathrm{imp}}$ by a solver of choice and interactively search for $\mu$ according to eq \ref{imp_Hamil} and \ref{no_elec_opt}. 
\item Solve for $u_{pq}$ according to eqs \ref{mf_Hamil}-\ref{CF:1}. If not converged, return to step 5.
\item If $u_{pq}$ is converged, compute the total energy per cell and the band structure using eq \ref{CF:2}.
\end{enumerate}

\begin{figure}[H]
    \centering
    \scalebox{1.0}{\includegraphics{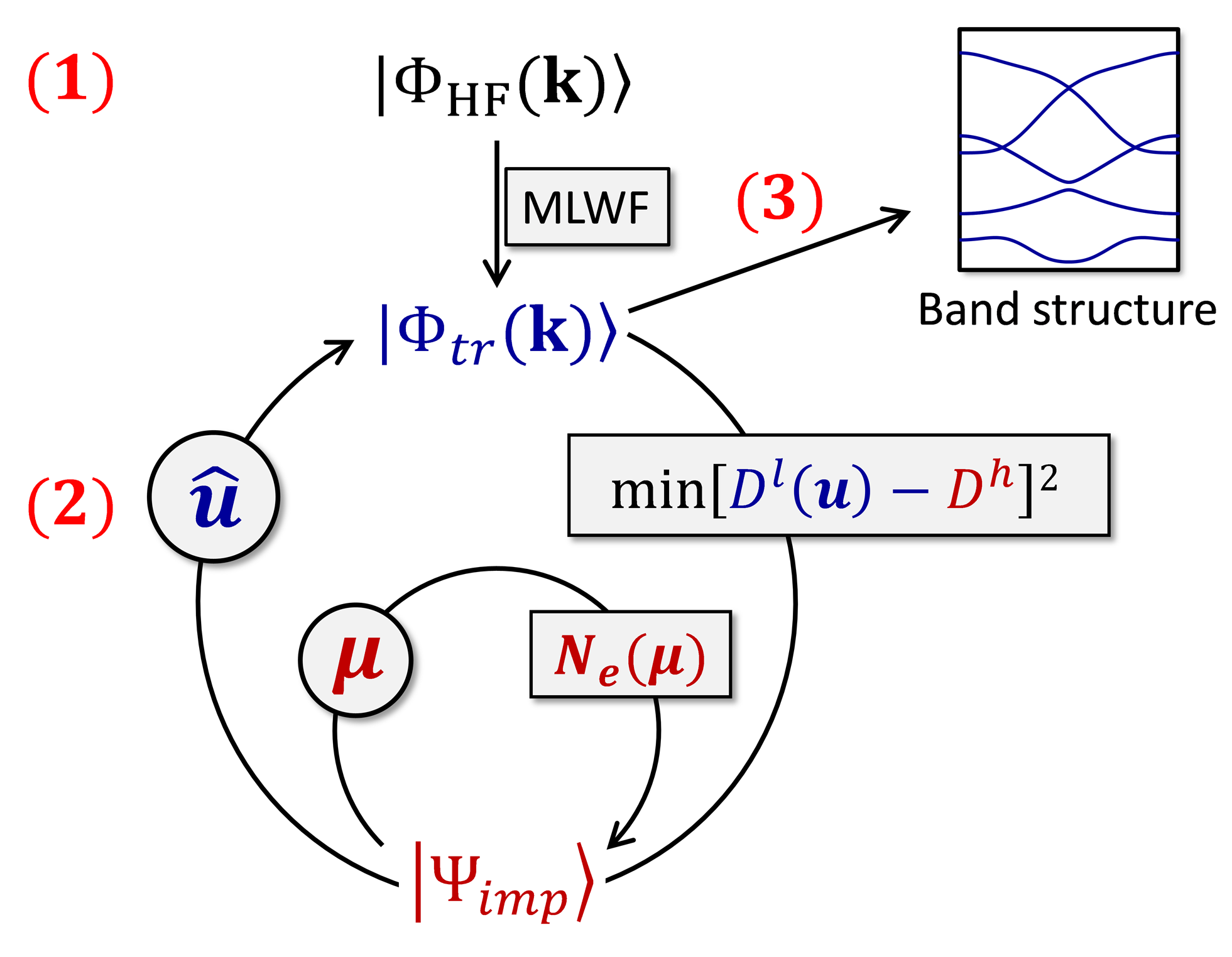}}
    \caption{Schematic diagram for the perodic DMET algorithm: (1) Construct the HF wave function and MLWFs, (2) Perform DMET interactive cycles, (3) Compute the band structure.}
    \label{pDMET_algorithm}
\end{figure}

As mentioned above, we exploit the dual representation of periodic systems in our implementation where the bath updating is performed in the \textbf{k}-space while the impurity problem is solved in the real space. Currently, the ERI transformation in step 3 requires substantial memory and computational time and scales poorly with the number of \textbf{k}-points. Fortunately, this step is performed only once and is parallel-scalable using OpenMP or MPI techniques.\cite{parallel} The second factor adding to the computational cost of the periodic DMET is the self-consistency in steps 4-9. Although the cost for a single impurity problem is manageable, the convergence of the correlation potential is often slow when $\hat{u}$ contains many degrees of freedom. This convergence problem arises in the conventional DMET algorithm. Although there have been several efforts to improve the convergence of the DMET algorithm,\cite{BE1D,p-DMET} a definite answer has not yet been found. In the current work, we either do a fully self-consistent DMET (sc-DMET) or a one-shot DMET (o-DMET), which only includes step 1-8. This is  similar to different flavors in a GW calculation, which depends on how one would like to optimize the self-energy.\cite{GW_VASP1,GW_VASP2,GW_VASP3} The periodic DMET algorithm is presented by the schematic diagram in Figure \ref{pDMET_algorithm}. 
The primary difference between the periodic DMET algorithm and the molecular one is the exploitation of translational symmetry. In fact, one can apply the molecular DMET algorithm to a supercell of a crystal (the size can be made equal to the size of any computational supercell defined in the Section 2.1) starting with a mean-field wave function at the $\Gamma$-point. The TDL convergence of the total energy by using a large supercell should be equivalent to that by sampling the FBZ with a dense \textbf{k}-mesh. However, the detail of the wave function in the \textbf{k}-space would be lost in the former case and the construction of a band structure would not be possible. Our algorithm strictly preserves translational invariance by choosing one unit cell as the fragment as well as by exploiting the \textbf{k}-space representation of the Hamiltonian. 
Furthermore, our use of MLWFs in DMET has several advantages. As previously mentioned, the construction of MLWFs can be performed on top of any band structure methods (\textit{e.g.}, Gaussian, plane-wave, numerical basis set) thereby facilitating the interface of DMET with a periodic mean-field code. Secondly, the construction of a smooth band structure requires a large number of \textbf{k}-points, hence it becomes intractable for expensive band structure methods, such as KS-DFT using exact exchange, GW, or equation-of-motion coupled cluster singles and doubles (EOM-CCSD) method.\cite{kEOM-CCSD} Meanwhile, the DMET band structure can be conveniently constructed from a smaller \textbf{k}-point mesh calculation (but large enough to reach the TDL) using the MLWF-based interpolation scheme.\cite{wannier90} In fact, this interpolation has been implemented for hybrid functionals and GW band structures in the VASP package.\cite{VASP1,VASP2,VASP3,VASP4}

\section{Computational methods \label{sec_comput_details}}
\begin{figure}
    \centering
    \scalebox{.6}{\includegraphics{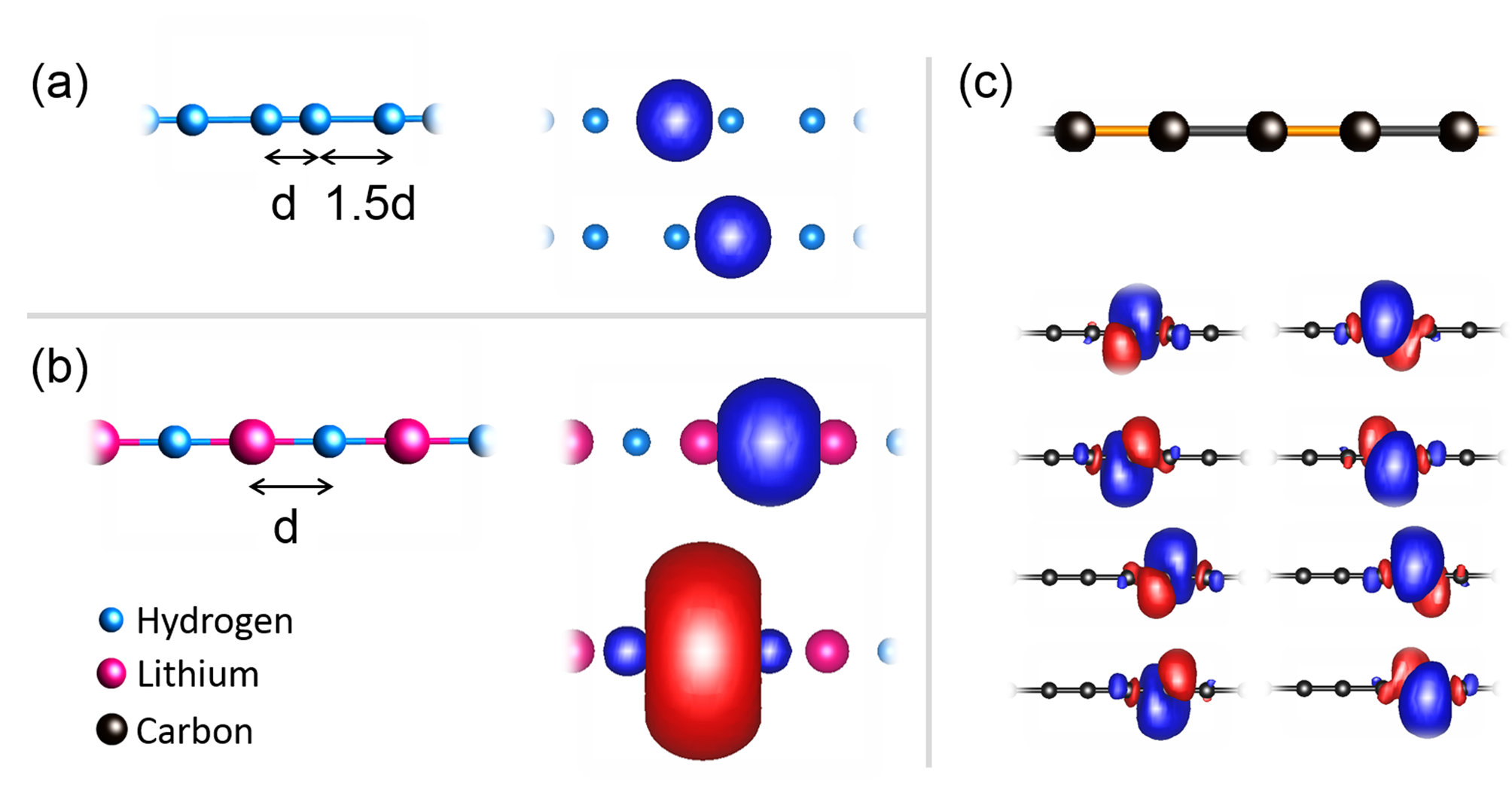}}
    \caption{Structure and MLWFs for (a) 1D-H  (b) 1D-LiH  (c) Polyyne}
    \label{testing_systems}
\end{figure}

In order to test the performance of periodic DMET and our implementation, we first compute the total energy as a function of bond length (or lattice parameter) for several one-dimensional solids: hydrogen (1D-H), lithium hydride (1D-LiH), and polyyne. All systems share a similar crystal structure with two atoms per unit cell. For 1D-H, the internuclear distance between adjacent hydrogen atoms alternates between two lengths with a ratio between them of 1.5 (Figure \ref{testing_systems}a).  For 1D-LiH, all adjacent Li-H internuclear distances are set to the same value (Figure \ref{testing_systems}b). For polyyne, the length for the single (grey) and triple (orange) bonds at equilibrium (scaling factor = 1) are 1.320 $\textrm{\AA}$ and 1.263 $\textrm{\AA}$, respectively (Figure \ref{testing_systems}c).  These systems are quasi-1D in the sense that they are constructed by separating adjacent chains by a distance of  10 $\textrm{\AA}$. We use GTH pseudopotentials\cite{GTH} and the SZV basis set for all the atoms. This corresponds to one, two, and four orbitals for H, Li, and C, respectively. For all structures, the FBZ is sampled by a uniform $\Gamma$-centered \textbf{k}-mesh. The total energy obtained from DMET is compared against the exact solution using the molecular solvers on the computational supercell at the same level of theory (FCI or CCSD). \\

Next, we benchmark the quasiparticle band structures from DMET against EOM-CCSD\cite{kEOM-CCSD} as well as KS-DFT using the PBE\cite{PBE} and B3PW91\cite{B3PW91a,B3PW91b} functionals. The computed band structures are compared to results from IP-EOM-CCSD and EA-EOM-CCSD.\cite{kEOM-CCSD} For each case, we seek for one root which has the largest overlap with the single excitation. These Koopmans-like states are good approximations for the conduction ($\equiv - EA(\textbf{k})$ where $EA(\textbf{k})$ is the EA at the crystal momentum $\textbf{k}$) and valence band ($\equiv -IP(\textbf{k})$ where $IP(\textbf{k})$ is the IP at the crystal momentum $\textbf{k}$). \\

DMET calculations are performed using our code, called pDMET, which can be found in the github.\cite{pDMET} Note that our code is not limited to one-dimensional systems; 1D solids are used to test our implementation because the exact solution for them is affordable. The molecular quantum chemical solvers, the electron repulsion integrals (ERIs), periodic HF, periodic KS-DFT, and periodic EOM-CCSD are obtained using the PySCF 1.6 package.\cite{PySCF} The PBC module of PySCF utilizes a hybrid basis set scheme where the localized nature of Gaussian functions and the efficiency of plane-waves are combined to treat systems under PBC. For more details, we refer the readers to existing papers.\cite{kEOM-CCSD} The MLWFs are constructed using the wannier90 code\cite{wannier90} via our Python interface pyWannier90.\cite{pyWannier90} All the calculations are performed in the spin restricted formalism.

\section{Results and discussions \label{sec_results}}
\subsection{Ground-state energy \label{sec_results_gse}}

\begin{figure}[H]
    \centering
    \includegraphics{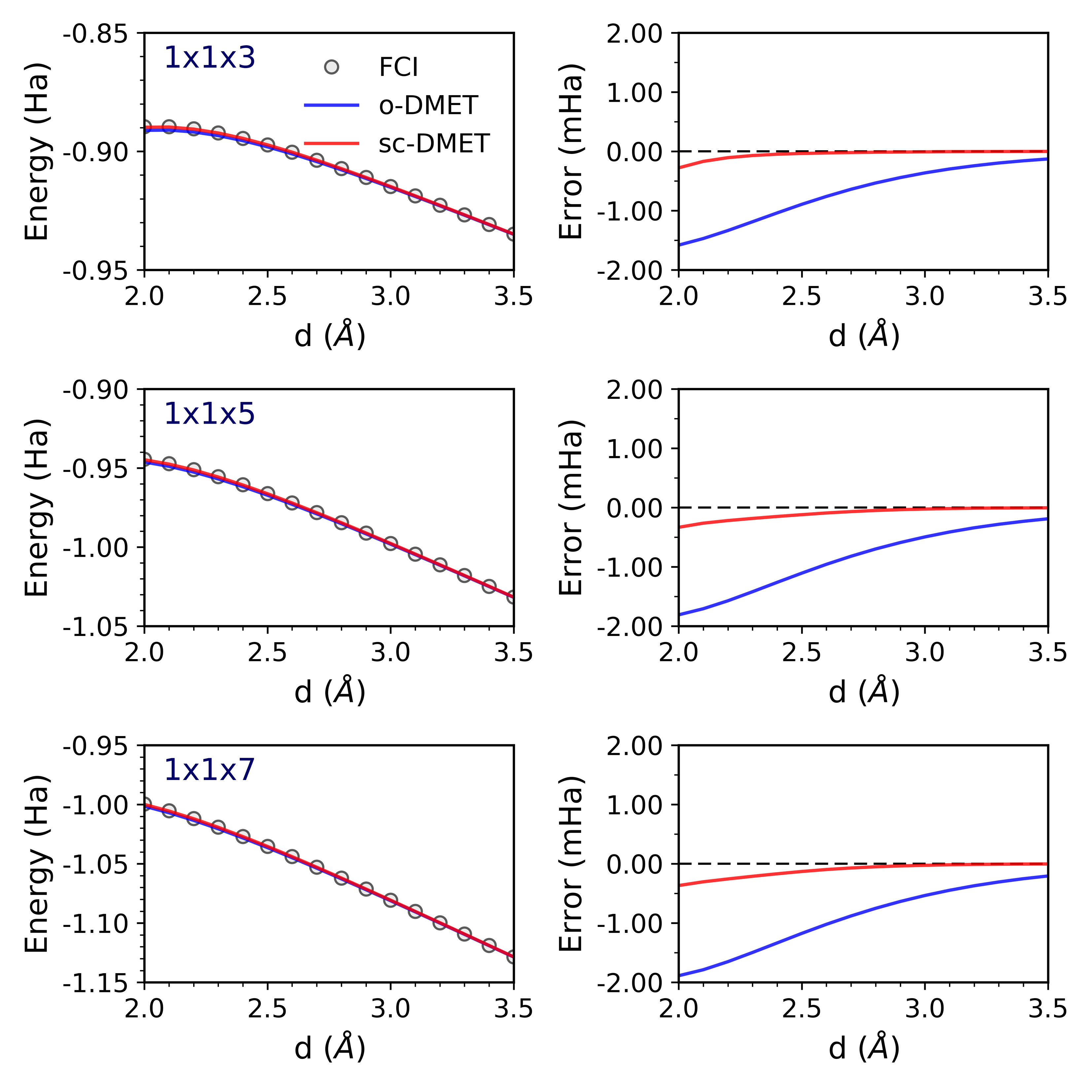}
    \caption{Left panels: Total energy per unit cell for 1D-H as a function of the bond length computed by different methods: one-shot DMET (o-DMET), self-consistent DMET (sc-DMET), and FCI; Right panels: the error  with respect to the FCI energy.}
    \label{1H_eos}
\end{figure}

\textbf{One-dimensional hydrogen (1D-H)}: One-dimensional hydrogen is a simple, but useful toy system exhibiting strong electron correlation. In this model system, electron correlation is stronger when the H-H distance is larger and weaker when it is smaller. In 1D-H, the band structure contains  two \textit{s}-bands, resulting in two \textit{s}-like MLWFs (Figure \ref{testing_systems}a). Figure \ref{1H_eos} presents the total energy using different numbers of \textbf{k}-points as a function of the distance (d). Our result shows that both o-DMET and sc-DMET (FCI is used as the high-level solver) give excellent agreement with the FCI energy with errors lower than 2 mHa. Remarkably, the accuracy of DMET is consistent with the number of \textbf{k}-points, indicating that DMET is \textit{de facto} size-extensive. This is an important requirement for an electronic structure theory for periodic systems as one often needs to converge the total energy to the TDL by using a dense \textbf{k}-mesh. Moreover, the sc-DMET energy is almost identical to the FCI energy, demonstrating the ability of the self-consistent optimization of the correlation potential to capture the electron correlation.

\begin{figure}[H]
    \centering
    \scalebox{1.0}{\includegraphics{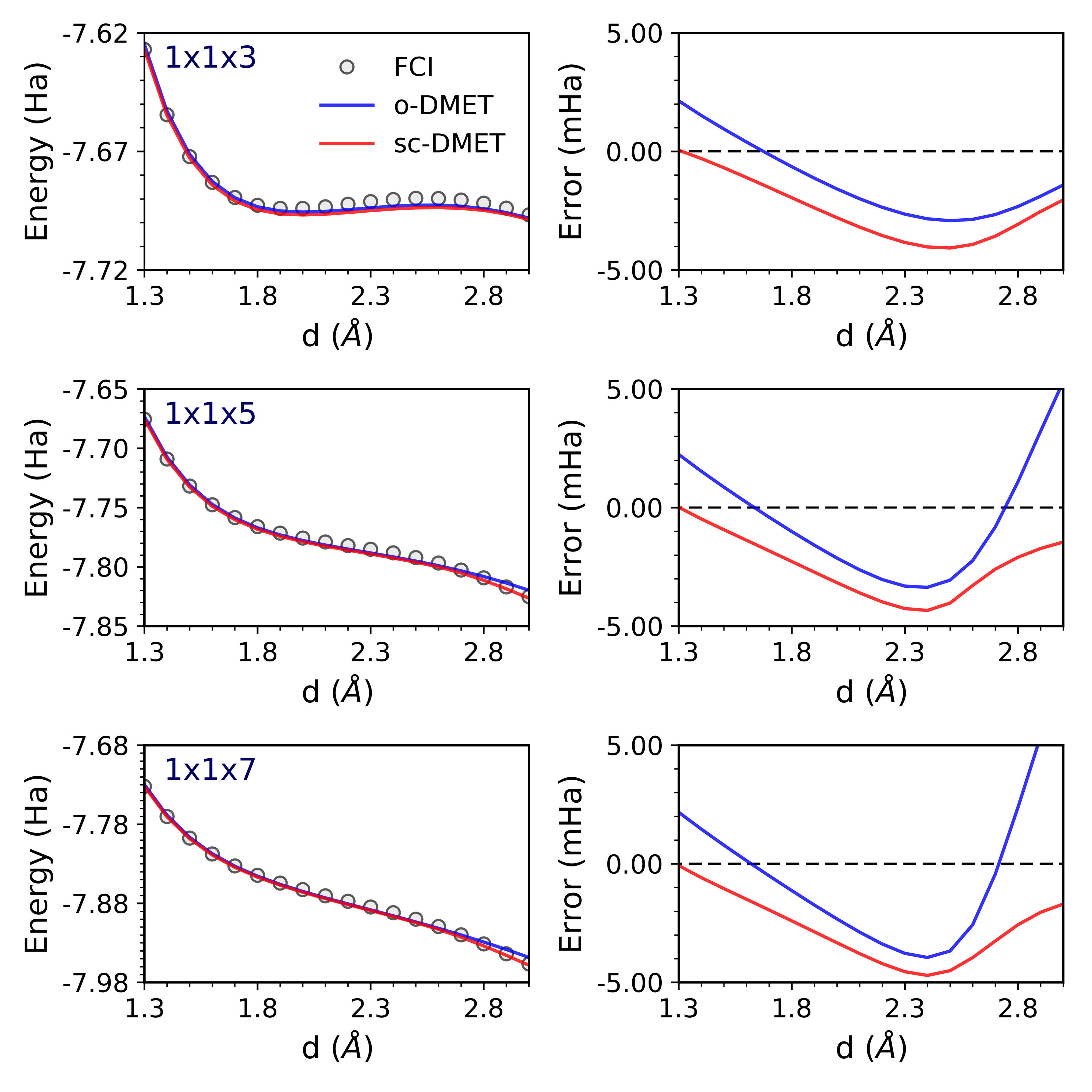}}
    \caption{Left panels: Total energy per unit cell for 1D-LiH as a function of the bond length computed by different methods: one-shot DMET (o-DMET), self-consistent DMET (sc-DMET), and FCI; Right panels: the error  with respect to the FCI energy.}
    \label{LiH_eos}
\end{figure}

\textbf{One-dimensional lithium hydride (1D-LiH)}: Similar to 1D-H, we performed the same calculations for 1D-LiH. For this system each unit cell contains three \textit{s}-orbitals, resulting in three well-separated \textit{s}-bands. Since the core Li 1\textit{s}-band is far from the Fermi level, it can be kept frozen in the DMET calculation. The other two bands are included in the constructions of two MLWFs which locate at the hydrogen and lithium atom (Figure \ref{testing_systems}b). As shown in Figure \ref{LiH_eos}, there is a small difference between o-DMET and sc-DMET when a 1x1x3 \textbf{k}-mesh is used. However, o-DMET with a denser \textbf{k}-mesh describes the dissociation limit poorly, exhibiting substantial and increasing undercorrelation for distances larger than 2.9 $\textrm{\AA}$. Meanwhile, the errors of sc-DMET are less than 5 mHa across the entire potential energy curve and again consistent between the different numbers of \textbf{k}-points.

\begin{figure}[H]
    \centering
    \scalebox{1.0}{\includegraphics{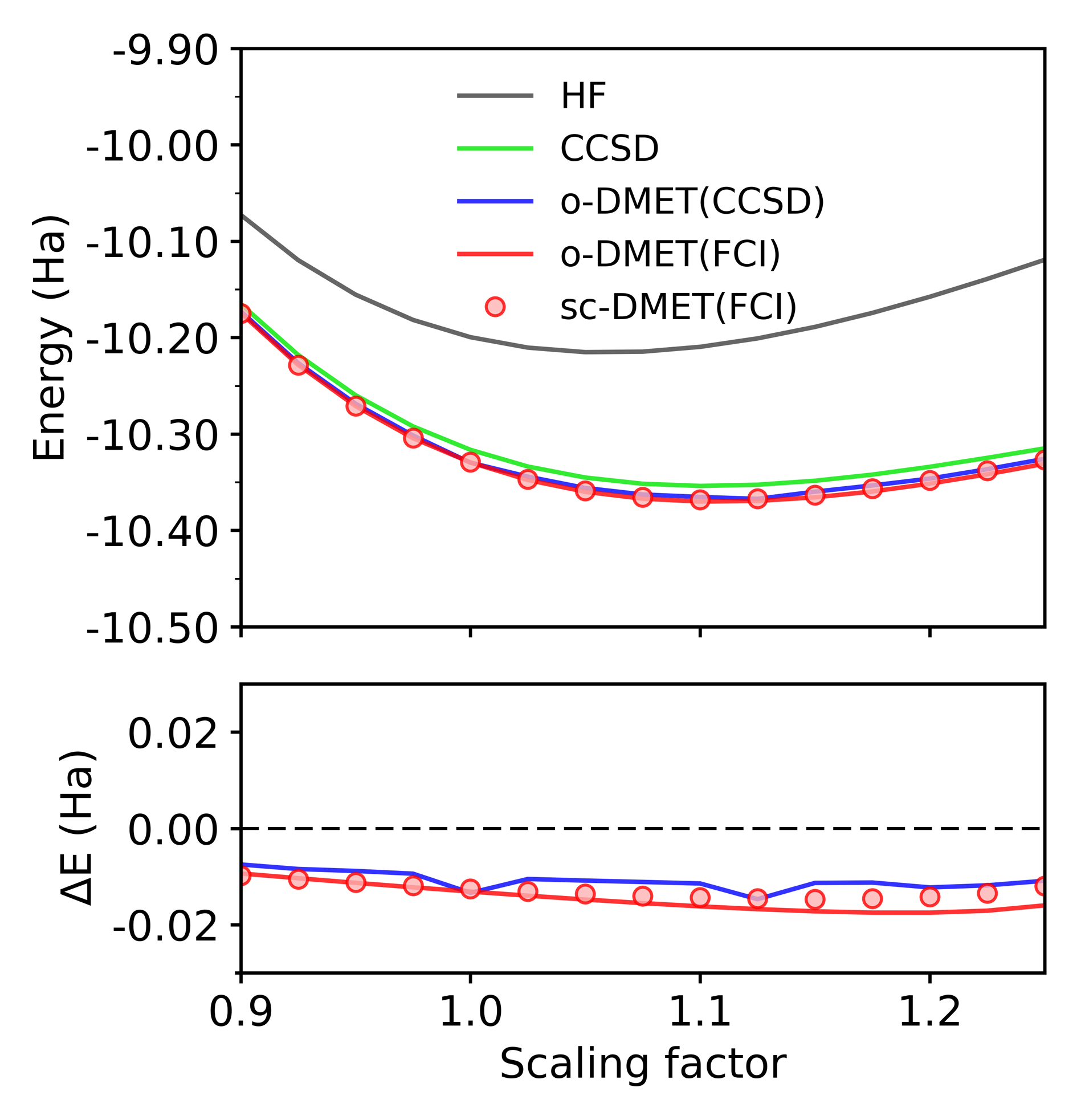}}
    \caption{Total energy per unit cell for polyyne as a function of the scaling factor. $\Delta$E is the difference between the CCSD energy and that of DMET.}
    \label{polyyne_eos}
\end{figure}

\textbf{Polyyne}: All eight bands are used to generate MLWFs from a HF wave function using 3 \textbf{k}-points. As a result, the total system has 24 electrons in 24 orbitals while the embedding space is only 16 electrons in 16 orbitals. We perform o-DMET using CCSD and FCI as well as sc-DMET using FCI as the high-level solvers. The difference between o-DMET(CCSD) and CCSD is less than 0.01 Ha at all geometries (Figure \ref{polyyne_eos}). There is a small difference between o-DMET(CCSD) and o-DMET(FCI) for the scaling factor smaller than 1. However, the difference becomes larger (up to \textit{ca}. 9 mHa) when the system is stretched (the scaling factor larger than 1). In general, the o-DMET(CCSD) and o-DMET(FCI) dissociation curves qualitatively agree with each other around, implying that the electron correlation in polyyne is weak, particularly around the equilibrium geometry. We note that the energies by sc-DMET(FCI) are slightly larger than those by o-DMET(FCI), the largest difference being only \textit{ca.} 2 mHa.

Our o-DMET(CCSD) calculations are essentially equivalent to the ``DET'' calculations reported in the left panel of fig 1 of Ref. \citenum{DET_solid}, except that we use a different AO basis set. With that in mind, the o-DMET(FCI) and sc-DMET(FCI) results show that for this system, the use of FCI as the solver and the optimization of the bath orbitals do not substantially affect the difference between DMET and CCSD total energies, implying that the DMET formalism itself is responsible for the disagreement.

\subsection{Electronic band structure \label{sec_results_band}}
We compute quasiparticle band structures for the 1D-H (d = 2.0 $\textrm{\AA}$) and polyyne (equilibrium, scaling factor = 1.0) using different levels of theory. As aforementioned, a smooth band structure from HF, KS-DFT, or DMET can be straightforwardly constructed by means of an interpolation scheme using MLWFs. In order to construct continuous band states from EOM-CCSD, a dense \textbf{k}-mesh must be shifted around to get the IP and EA at the desired \textbf{k}-point. In particularly, we first perform a EOM-CCSD calculation using a uniform mesh of 15 \textbf{k}-points centered at $\Gamma$ ($\textbf{k} = 0$) to sample the FBZ.  We then shifted the same \textbf{k}-mesh so that it centers at Z ($\textbf{k} = 0.5$) and performed another calculation using this shifted \textbf{k}-mesh while keeping the other parameters unchanged. This procedure is only valid if the wave functions using different \textbf{k}-meshes converge to the same electronic state. Indeed, the difference in the CCSD energies are less than 0.01 mHa for both structures (see Table 8 in the \textbf{SI}), hence the EOM-CCSD energies from two \textbf{k}-meshes can be combined to construct a smooth conduction and valence band. 

\begin{table}[H]
\begin{center}
\caption{Band gap  for 1D-H and polyyne from different methods.}
\begin{tabular}{*4c}    \toprule
& & \multicolumn{2}{c}{Band gap (eV)} \\  \cmidrule(){3-4}
& \multicolumn{1}{l}{Method} & This work &  Literature\\ \toprule
1D-H   & \multicolumn{1}{l}{HF}  & 14.53 & \\ 
             & \multicolumn{1}{l}{PBE}  & 2.00 & \\
             & \multicolumn{1}{l}{B3PW91}  & 4.48 & \\
             & \multicolumn{1}{l}{EOM-CCSD}  & 20.46 & \\
             & \multicolumn{1}{l}{DMET}  & 17.84 & \\ \midrule
             
Polyyne  & \multicolumn{1}{l}{HF}  & 5.88 & \\
             & \multicolumn{1}{l}{PBE}  & 0.58 &  \\
             & \multicolumn{1}{l}{B3PW91}  & 1.14 &  \\
             & \multicolumn{1}{l}{EOM-CCSD}  & 5.06 & \\
             & \multicolumn{1}{l}{DMET}  & 3.96 & \\ 
   			 & \multicolumn{1}{l}{DMC}  & & 3.61\cite{DMC_polyyne}\\ 
   			 & \multicolumn{1}{l}{GW}  & & 2.16\cite{GW_polyyne}\\ \toprule
\end{tabular}
\end{center}
\end{table}

\begin{figure}[H]
    \centering
    \scalebox{1.0}{\includegraphics{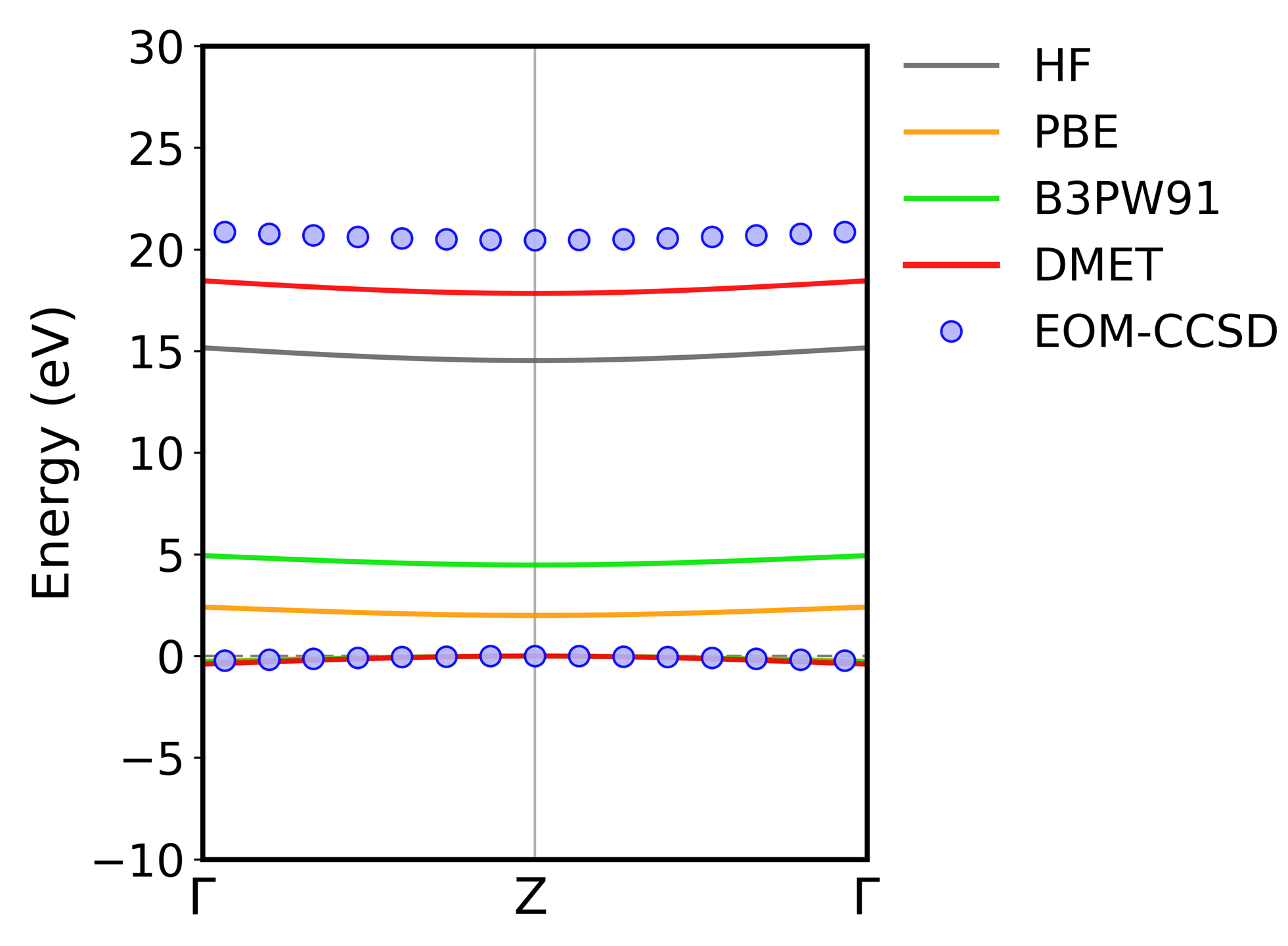}}
    \caption{Electronic band structure of hydrogen chain computed by sc-DMET using a CCSD solver and other methods. Note that the valence bands by different methods are almost identical. The Fermi energy is shifted to 0 eV.}
    \label{1H_band}
\end{figure}

\textbf{1D-H}: Figure \ref{1H_band} shows the band structures for the 1D-H (d = 2.0 $\textrm{\AA}$) using different levels of theory. The bands from different theories have similar dispersion and the difference here is mainly in the band gap. The band gaps from KS-DFT using either PBE (2.00 eV) or B3PW91  (4.48 eV) functionals are much smaller than the HF band gap (14.53 eV) (see Table 1). Interestingly, sc-DMET with CCSD solver and EOM-CCSD  give higher band gaps of 17.84 and 20.46 eV, respectively. It is not feasible to conclude which method is more accurate for this toy system due to the lack of a reference value. We note that although EOM-CCSD has been shown to accurately describe the band structures of diamond and silicon,\cite{kEOM-CCSD} its performance in predicting band gap for solids requires an extensive benchmark on a large number of systems.

\begin{figure}[H]
    \centering
    \scalebox{1.0}{\includegraphics{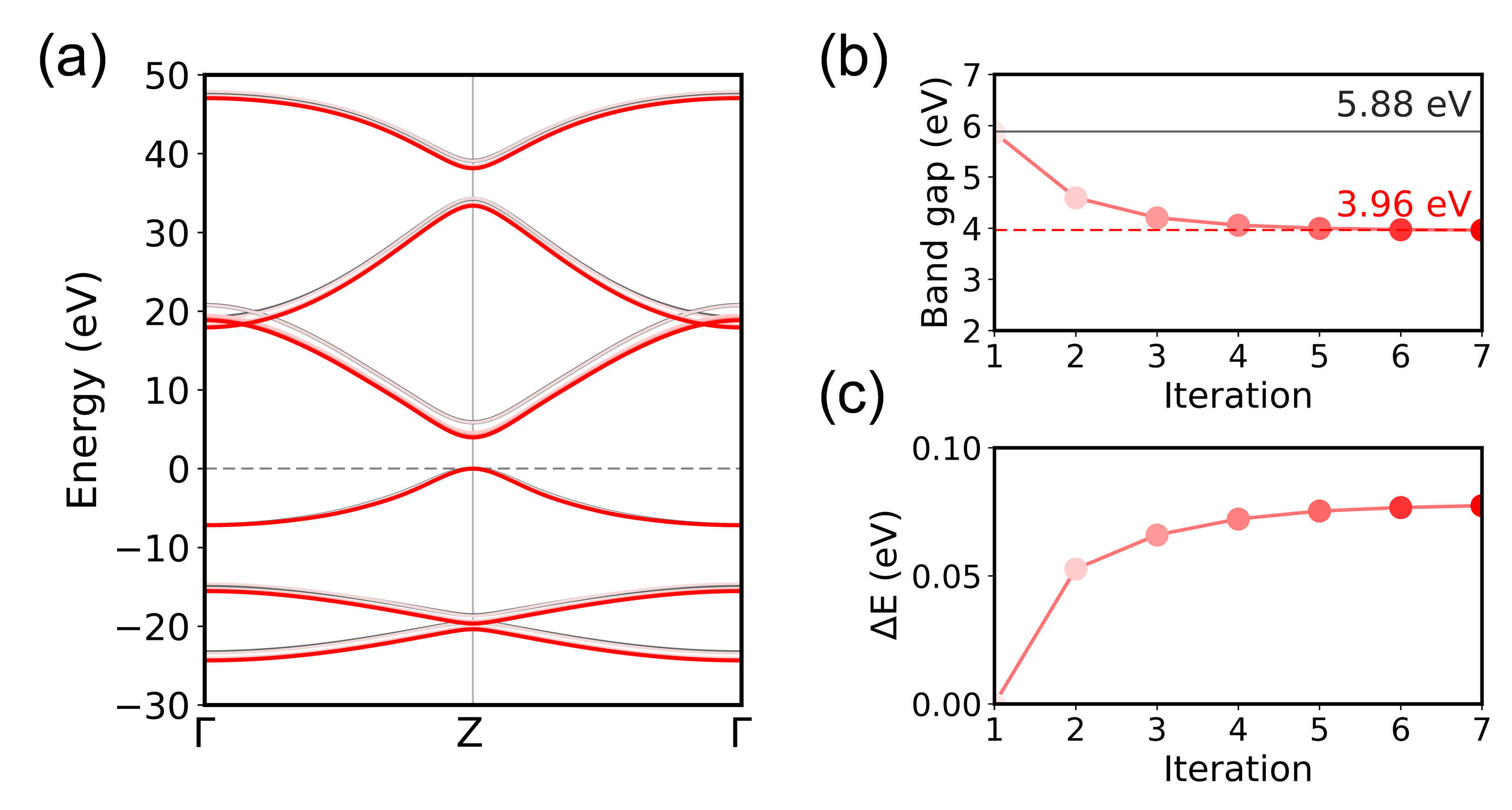}}
    \caption{Electronic band structure of polyyne computed by HF (grey) and DMET using the FCI solver (red) with the Fermi energy shifted to 0 eV (a), the convergence of band gap (b) and the total energy (c) against the number of iterations. $\Delta$E is the difference between the total energy at the n-th iteration and the first iteration (\textit{i.e.}, o-DMET.)}
    \label{polyyne_band:1}
\end{figure}

\begin{figure}[H]
    \centering
    \scalebox{1.0}{\includegraphics{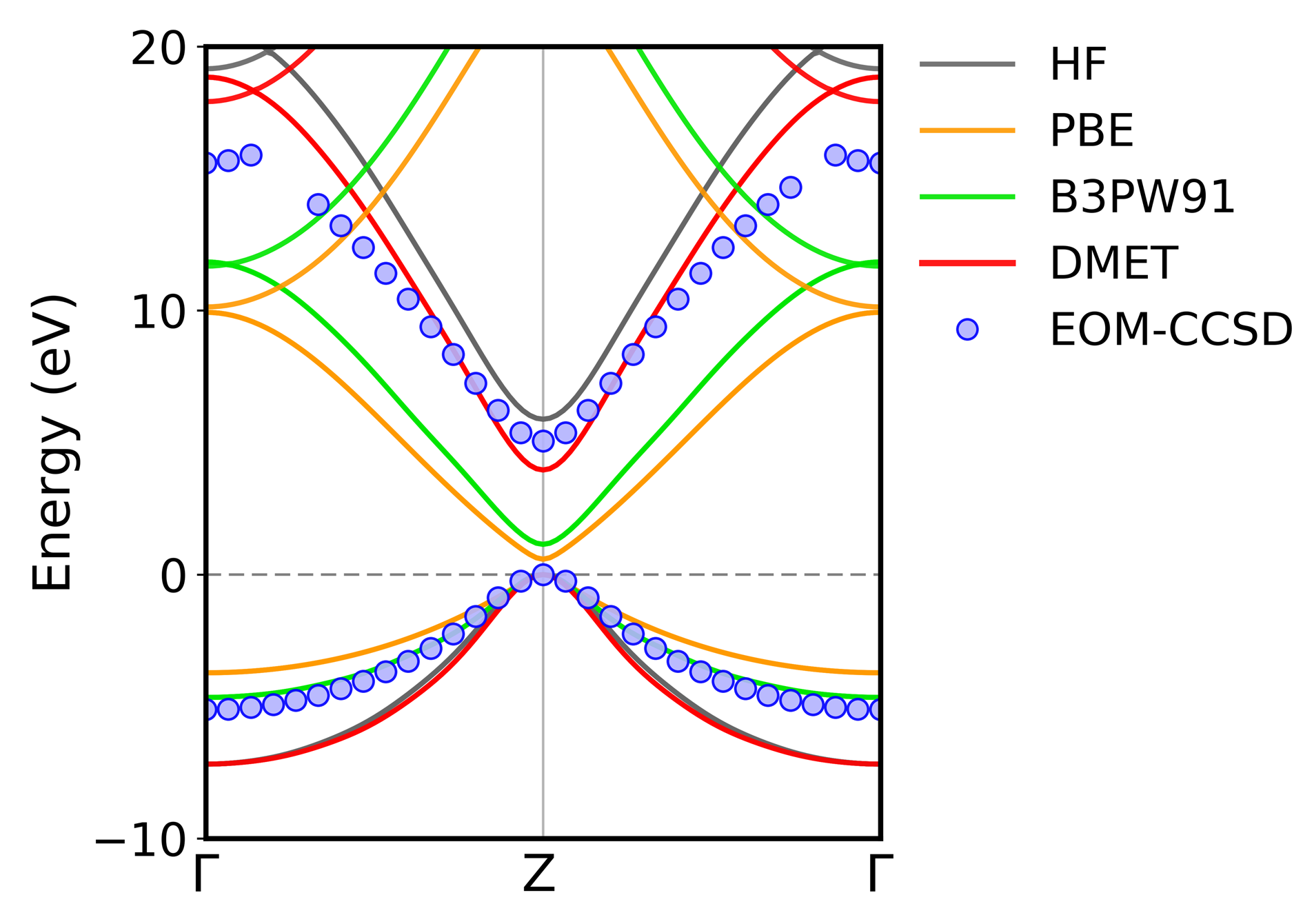}}
    \caption{Electronic band structure of computed computed by DMET (7 iterations) using a FCI solver and other methods. The Fermi energy is shifted to 0 eV.}
    \label{polyyne_band:2}
\end{figure}

\textbf{Polyyne}: The total energy from sc-DMET using CCSD solver does not converge to a stationary point after our exhausted effort. However, DMET using FCI quickly converges after several iterations within a threshold of 0.01 eV in band gap as shown in Figure \ref{polyyne_band:1}b. The difference in the total energy between FCI and CCSD is small at equilibrium geometry, as discussed in the previous section. Therefore, the band structure by the FCI solver can be compared directly with the EOM-CCSD one. The change in band gap after one iteration (\textit{i.e.}, o-DMET) is unnoticeable (although not identically zero because in this case we are still solving Eq. (\ref{CF:2}) for $\hat{u}^{'}$, even though $\hat{u}$ is zero). However, it is drastically decreasing and converged to an optimal value of 3.96 eV after 7 iterations, to be compared with the HF band gap of 5.88 eV (Figure 9b). Interestingly, the total energy difference between o-DMET and sc-DMET is only about 0.07 eV, which is much smaller than the change in the band gap. In general, the band dispersion described by HF and DMET are very similar (see Figure \ref{polyyne_band:1}a).
Next, we compare the DMET band structure, obtained using FCI as a solver, against other theories (see Figure \ref{polyyne_band:2}). Generally speaking, HF and DMET bands are the most dispersive ones while PBE predicts the least dispersive bands amongst all the methods. The EOM-CCSD bands are as dispersive as the B3PW91 bands as one can see that their valence bands are nearly on top of each other. The conduction and valence bands from DMET and EOM-CCSD mainly differ at the high-symmetric points (\textit{i.e.}, $\Gamma$ and Z), while in the intermediate region there is good agreement between the two methods. Band structure of materials can be experimentally studied  by angle-resolved photoemission spectroscopy (ARPES)\cite{ARPES_visualizing,ARPES_quantummat,ARPES_cuprate}. Unfortunately, ARPES measurements are not available for polyyne, hence it is not possible to conclude further which theory is the most accurate in terms of describing the band dispersion.

Similar to 1D-H, KS-DFT predicts a small band gap for polyyne; in particular, the PBE and B3PW91 band gaps are 0.58 eV and 1.14 eV, respectively, which are much smaller than the HF gap (5.88 eV). The EOM-CCSD band gap (5.06 eV) is slightly smaller than that of HF. Meanwhile, the DMET band gap is of 3.96 eV, which is \textit{ca.} 2 eV smaller than the HF band gap. The experimental determination of the band gap for an infinite linear chain of carbon is very challenging due to its extreme instability. Although the band gap for some short chains of polyyne (up to 44 carbon atoms) have been determined by different techniques, a consensus has not been reached yet in the literature.  Therefore, to resolve the disparity between DMET and EOM-CCSD we also compare our calculations with other many-body techniques from the literature. The DMET band gap is very close to the quasiparticle gap of 3.61 eV computed by diffusion quantum Monte Carlo (DMC),\cite{DMC_polyyne} and somewhat close to that of 2.16 eV computed by GW calculations.\cite{GW_polyyne} This is a substantially better agreement than that of EOM-CCSD at 5.06 eV, which is an encouraging sign about the accuracy of DMET's band structures as well as its overall accuracy in predicting solid-state band gaps.

\section{Concluding remarks \label{sec_conclusion}}

We have introduced and implemented periodic DMET as a \textit{ab initio} band structure method based on density matrix embedding for periodic systems. By exploiting crystal translations, DMET can provide a quasiparticle band picture for solid-state materials. DMET shares the same core idea as DFT+DMFT in which the local electron correlation can be utilized to improve the global mean-field observables. Unlike DFT+DMFT, DMET by construction is free from double-counting issues and computationally less expensive owing to its frequency-independent formulation. In principle, with DMET one can compute the entire electronic band structure which is very important in studying the non-trivial topological structure of materials.\cite{TQC} Moreover, we emphasize that periodic DMET is not yet designed to study local observables in system with broken (short-range) translational symmetry, for example, a crystal defect or an catalytic active site or gas molecules on a surface. For these systems, other quantum embedding techniques, such as density functional embedding theory\cite{PDFET_CASSCF}, projection-based embedding theory,\cite{ProEmb} LASSCF,\cite{LASSCF} active-space decomposition method (ASD),\cite{ASD,ASD_DMRG,ASD_SCF} or molecular DMET can be used.

Our preliminary results on one-dimensional solids are promising. The efficiency of DMET is dominated by two processes: the integral transformation and the convergence of the correlation potential. While the former can be improved by exploring the parallel computing techniques, the latter (also arises in the molecular DMET) is more problematic. Furthermore, the convergence of the correlation energy toward the thermodynamic limit has not yet been inspected in this work. These issues, currently under investigation, are essential to make DMET a practical method.  Finally, DMET can progress in several directions. For example, a real-time extension of DMET or dynamical cluster approximation formulation can be developed in analogy to those of DMET for molecule and lattice models.\cite{DMET_realtime,DCA-DMET} We also plan to explore difference solvers in combination with periodic DMET like multireference methods\cite{MC_methods} to study band structures. We anticipate that our proposed theory can be an potential alternative to DFT+DMFT as a band structure method for strongly correlated materials. 
    
\section*{Acronyms}

\begin{table}[H]
\begin{tabular}{K{3cm} K{10cm}}   
\toprule
ASD &  Active-space decomposition\\
CCSD & Coupled cluster singles and doubles\\
CF & Cost function\\
DCA & Dynamical cluster approximation\\
DFT &  Discrete Fourier transform\\
DET & Density embedding theory\\
DMC & Diffusion quantum Monte Carlo\\
DMET & Density matrix embedding theory\\
o-DMET & One-shot DMET\\
sc-DMET & Self-consistent DMET\\
DMFT & Dynamical mean-field theory\\
EOM-CCSD & Equation of motion coupled cluster singles and doubles\\
FBZ & First Brillouin zone\\
FCI	&  Full configuration interaction\\
HF & Hartree-Fock\\
GGA & Generalized gradient approximation\\
KS-DFT & Kohn-Sham density functional theory\\
LASSCF & Localized active space self-consistent field\\
LDA & Local density approximation\\
MLWF & Maximally-localised Wannier function\\
SCF & Self-consistent field\\
TDL & Thermodynamic limit \\
\toprule
\end{tabular}
\end{table}

\section*{Associated content}
\textbf{Supporting Information}. Absolute energies at different geometries for 1D-H, 1D-LiH, and polyyne are reported. This material is available free of charge via the Internet at http://pubs.acs.org.

\section*{Associated content}
\textbf{Corresponding Author} \\
*gagliard@umn.edu (L.G.) \\
\textbf{Author Contributions} \\
The manuscript was written through contributions of all authors. All authors have given approval to the final version of the manuscript. \\
\textbf{Notes} \\
The authors declare no competing financial interests.

\acknowledgement
We thank Christopher J. Cramer, Riddhish U. Pandharkar, and Donald G. Truhlar for insightful discussion. H. Q. P. thanks Qiming Sun, Garnet K.-L. Chan, and Timothy C. Berkelbach for helpful support on PySCF. This research is supported by the U.S. Department of Energy, Office of Basic  Energy Sciences, Division of Chemical Sciences, Geosciences and Biosciences under Award DEFG02-17ER16362. Computer resources were provided by the Minnesota Super-computing Institute at the University of Minnesota.

\bibliography{ref}

\section*{Table of Contents graphic}

\begin{figure}[H]
    \centering
    \scalebox{1.0}{\includegraphics{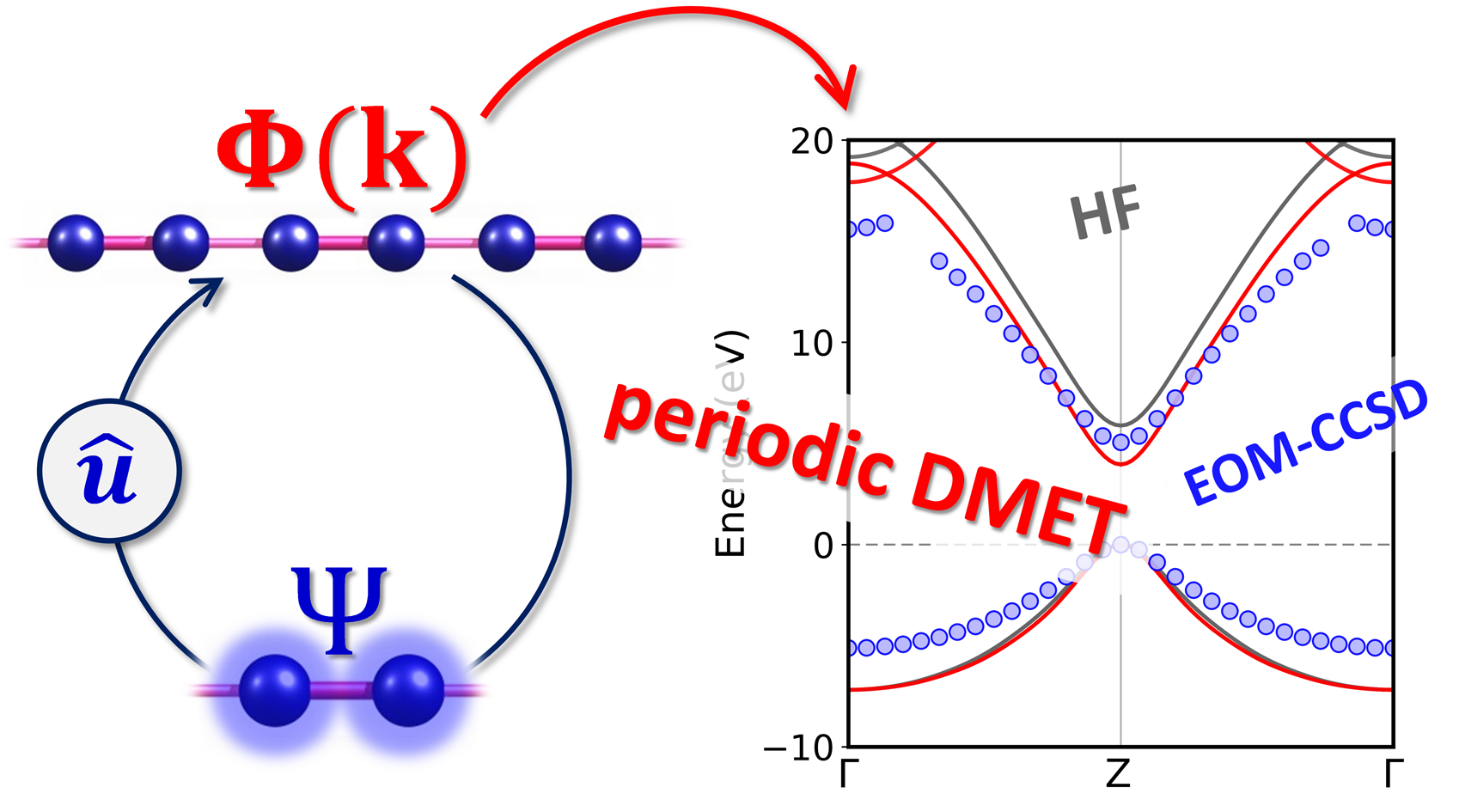}}
    \label{TOC}
\end{figure}

\end{document}


\tableofcontents
\clearpage

\section{Total energies for 1D-H}

\begin{table}[H]
\begin{center}
\caption{Total energy per unit cell (in Hartree) as a function of the H-H distance (d) for 1D-H using 3 \textbf{k}-points.}
\begin{tabular}{K{1cm} K{3.5cm} K{3.5cm} K{3.5cm} K{3.5cm}}    \toprule
\textbf{d}	&   HF	&      FCI	&   o-DMET & sc-DMET \\
\hline
2.0 & -0.772724 & -0.889568 & -0.891149 & -0.889846 \\
2.1 & -0.760445 & -0.889564 & -0.891034 & -0.889733 \\
2.2 & -0.749271 & -0.890493 & -0.891828 & -0.890599 \\
2.3 & -0.739181 & -0.892176 & -0.893363 & -0.892246 \\
2.4 & -0.730149 & -0.894456 & -0.895494 & -0.894504 \\
2.5 & -0.722135 & -0.897200 & -0.898093 & -0.897235 \\
2.6 & -0.715095 & -0.900300 & -0.901060 & -0.900327 \\
2.7 & -0.708980 & -0.903670 & -0.904309 & -0.903690 \\
2.8 & -0.703736 & -0.907243 & -0.907776 & -0.907258 \\
2.9 & -0.699305 & -0.910967 & -0.911409 & -0.910978 \\
3.0 & -0.695629 & -0.914804 & -0.915167 & -0.914812 \\
3.1 & -0.692650 & -0.918724 & -0.919021 & -0.918729 \\
3.2 & -0.690307 & -0.922704 & -0.922947 & -0.922708 \\
3.3 & -0.688544 & -0.926730 & -0.926926 & -0.926733 \\
3.4 & -0.687304 & -0.930788 & -0.930947 & -0.930790 \\
3.5 & -0.686536 & -0.934871 & -0.934999 & -0.934872 \\ \toprule
\end{tabular}
\end{center}
\end{table}

\begin{table}[H]
\begin{center}
\caption{Total energy per unit cell (in Hartree) as a function of the H-H distance (d) for 1D-H using 5 \textbf{k}-points.}
\begin{tabular}{K{1cm} K{3.5cm} K{3.5cm} K{3.5cm} K{3.5cm}}    \toprule
\textbf{d}	&   HF	&      FCI	&   o-DMET & sc-DMET \\
\hline
2.0 & -0.825991 & -0.944402 & -0.946211 & -0.944735 \\
2.1 & -0.816455 & -0.947229 & -0.948935 & -0.947492 \\
2.2 & -0.808000 & -0.950976 & -0.952548 & -0.951195 \\
2.3 & -0.800618 & -0.955469 & -0.956888 & -0.955652 \\
2.4 & -0.794282 & -0.960551 & -0.961812 & -0.960701 \\
2.5 & -0.788959 & -0.966092 & -0.967198 & -0.966212 \\
2.6 & -0.784606 & -0.971986 & -0.972943 & -0.972077 \\
2.7 & -0.781175 & -0.978146 & -0.978967 & -0.978214 \\
2.8 & -0.778614 & -0.984506 & -0.985204 & -0.984555 \\
2.9 & -0.776866 & -0.991015 & -0.991605 & -0.991050 \\
3.0 & -0.775872 & -0.997635 & -0.998130 & -0.997660 \\
3.1 & -0.775575 & -1.004337 & -1.004749 & -1.004354 \\
3.2 & -0.775915 & -1.011099 & -1.011440 & -1.011110 \\
3.3 & -0.776835 & -1.017904 & -1.018186 & -1.017912 \\
3.4 & -0.778279 & -1.024741 & -1.024973 & -1.024747 \\
3.5 & -0.780194 & -1.031602 & -1.031791 & -1.031606 \\ \toprule
\end{tabular}
\end{center}
\end{table}

\begin{table}[H]
\begin{center}
\caption{Total energy per unit cell (in Hartree) as a function of the H-H distance (d) for 1D-H using 7 \textbf{k}-points.}
\begin{tabular}{K{1cm} K{3.5cm} K{3.5cm} K{3.5cm} K{3.5cm}}    \toprule
\textbf{d}	&   HF	&      FCI	&   o-DMET & sc-DMET \\
\hline
2.0 & -0.880530 & -0.999585 & -1.001473 & -0.999952 \\
2.1 & -0.873735 & -1.005201 & -1.006988 & -1.005505 \\
2.2 & -0.868019 & -1.011736 & -1.013387 & -1.011992 \\
2.3 & -0.863373 & -1.019014 & -1.020510 & -1.019225 \\
2.4 & -0.859773 & -1.026879 & -1.028214 & -1.027048 \\
2.5 & -0.857184 & -1.035202 & -1.036376 & -1.035332 \\
2.6 & -0.855564 & -1.043875 & -1.044896 & -1.043972 \\
2.7 & -0.854867 & -1.052813 & -1.053693 & -1.052884 \\
2.8 & -0.855038 & -1.061951 & -1.062702 & -1.062002 \\
2.9 & -0.856022 & -1.071237 & -1.071873 & -1.071273 \\
3.0 & -0.857762 & -1.080633 & -1.081168 & -1.080659 \\
3.1 & -0.860197 & -1.090111 & -1.090558 & -1.090128 \\
3.2 & -0.863270 & -1.099648 & -1.100019 & -1.099659 \\
3.3 & -0.866923 & -1.109228 & -1.109535 & -1.109236 \\
3.4 & -0.871100 & -1.118840 & -1.119092 & -1.118845 \\
3.5 & -0.875748 & -1.128475 & -1.128681 & -1.128478 \\ \toprule
\end{tabular}
\end{center}
\end{table}

\section{Total energies for 1D-LiH}

\begin{table}[H]
\begin{center}
\caption{Total energy per unit cell (in Hartree) as a function of the Li-H distance (d) for 1D-LiH using 3 \textbf{k}-points.}
\begin{tabular}{K{1cm} K{3.5cm} K{3.5cm} K{3.5cm} K{3.5cm}}    \toprule
\textbf{d}	&   HF	&      FCI	&   o-DMET & sc-DMET \\
\hline
1.0 & -7.421519 & -7.429404 & -7.424294 & -7.429658 \\
1.1 & -7.515142 & -7.522297 & -7.518575 & -7.522011 \\
1.2 & -7.578219 & -7.585117 & -7.582276 & -7.584829 \\
1.3 & -7.620026 & -7.627030 & -7.624895 & -7.626979 \\
1.4 & -7.647165 & -7.654537 & -7.653020 & -7.654838 \\
1.5 & -7.664227 & -7.672172 & -7.671228 & -7.672861 \\
1.6 & -7.674388 & -7.683078 & -7.682682 & -7.684178 \\
1.7 & -7.679829 & -7.689424 & -7.689557 & -7.690947 \\
1.8 & -7.682051 & -7.692713 & -7.693356 & -7.694665 \\
1.9 & -7.682092 & -7.693993 & -7.695122 & -7.696372 \\
2.0 & -7.680672 & -7.694008 & -7.695591 & -7.696803 \\
2.1 & -7.678300 & -7.693292 & -7.695289 & -7.696483 \\
2.2 & -7.675337 & -7.692240 & -7.694597 & -7.695788 \\
2.3 & -7.672040 & -7.691153 & -7.693799 & -7.694993 \\
2.4 & -7.668595 & -7.690263 & -7.693106 & -7.694293 \\
2.5 & -7.665136 & -7.689758 & -7.692681 & -7.693831 \\
2.6 & -7.661759 & -7.689784 & -7.692651 & -7.693710 \\
2.7 & -7.658532 & -7.690449 & -7.693114 & -7.694025 \\
2.8 & -7.655507 & -7.691819 & -7.694146 & -7.694896 \\
2.9 & -7.652718 & -7.693910 & -7.695802 & -7.696447 \\
3.0 & -7.650190 & -7.696691 & -7.698110 & -7.698739 \\ \toprule
\end{tabular}
\end{center}
\end{table}

\begin{table}[H]
\begin{center}
\caption{Total energy per unit cell (in Hartree) as a function of the Li-H distance (d) for 1D-LiH using 5 \textbf{k}-points.}
\begin{tabular}{K{1cm} K{3.5cm} K{3.5cm} K{3.5cm} K{3.5cm}}    \toprule
\textbf{d}	&   HF	&      FCI	&   o-DMET & sc-DMET \\
\hline
1.0 & -7.444103 & -7.451423 & -7.446105 & -7.448924 \\
1.1 & -7.548519 & -7.555075 & -7.551066 & -7.553635 \\
1.2 & -7.620186 & -7.626462 & -7.623425 & -7.625850 \\
1.3 & -7.669053 & -7.675403 & -7.673158 & -7.675386 \\
1.4 & -7.702219 & -7.708889 & -7.707354 & -7.709366 \\
1.5 & -7.724631 & -7.731803 & -7.730939 & -7.732735 \\
1.6 & -7.739705 & -7.747527 & -7.747307 & -7.748903 \\
1.7 & -7.749786 & -7.758391 & -7.758795 & -7.760214 \\
1.8 & -7.756481 & -7.766003 & -7.767010 & -7.768274 \\
1.9 & -7.760898 & -7.771481 & -7.773065 & -7.774202 \\
2.0 & -7.763805 & -7.775612 & -7.777740 & -7.778778 \\
2.1 & -7.765736 & -7.778964 & -7.781586 & -7.782558 \\
2.2 & -7.767070 & -7.781963 & -7.784997 & -7.785938 \\
2.3 & -7.768071 & -7.784948 & -7.788259 & -7.789206 \\
2.4 & -7.768928 & -7.788217 & -7.791581 & -7.792554 \\
2.5 & -7.769775 & -7.792064 & -7.795123 & -7.796090 \\
2.6 & -7.770707 & -7.796766 & -7.799004 & -7.800052 \\
2.7 & -7.771790 & -7.802498 & -7.803322 & -7.805095 \\
2.8 & -7.773074 & -7.809237 & -7.808159 & -7.811335 \\
2.9 & -7.774591 & -7.816799 & -7.813588 & -7.818526 \\
3.0 & -7.776364 & -7.824983 & -7.819682 & -7.826438 \\ \toprule
\end{tabular}
\end{center}
\end{table}

\begin{table}[H]
\begin{center}
\caption{Total energy per unit cell (in Hartree) as a function of the Li-H distance (d) for 1D-LiH using 7 \textbf{k}-points.}
\begin{tabular}{K{1cm} K{3.5cm} K{3.5cm} K{3.5cm} K{3.5cm}}    \toprule
\textbf{d}	&   HF	&      FCI	&   o-DMET & sc-DMET \\
\hline
1.0 & -7.486317 & -7.493478 & -7.488289 & -7.491069 \\
1.1 & -7.595887 & -7.602292 & -7.598385 & -7.600902 \\
1.2 & -7.672348 & -7.678484 & -7.675525 & -7.677937 \\
1.3 & -7.725805 & -7.732014 & -7.729838 & -7.732085 \\
1.4 & -7.763449 & -7.769967 & -7.768503 & -7.770549 \\
1.5 & -7.790285 & -7.797282 & -7.796495 & -7.798326 \\
1.6 & -7.809754 & -7.817369 & -7.817237 & -7.818863 \\
1.7 & -7.824218 & -7.832571 & -7.833077 & -7.834518 \\
1.8 & -7.835290 & -7.844499 & -7.845628 & -7.846903 \\
1.9 & -7.844080 & -7.854268 & -7.856005 & -7.857135 \\
2.0 & -7.851357 & -7.862662 & -7.864986 & -7.865992 \\
2.1 & -7.857656 & -7.870240 & -7.873120 & -7.874025 \\
2.2 & -7.863354 & -7.877418 & -7.880799 & -7.881626 \\
2.3 & -7.868717 & -7.884531 & -7.888306 & -7.889081 \\
2.4 & -7.873934 & -7.891896 & -7.895847 & -7.896603 \\
2.5 & -7.879140 & -7.899898 & -7.903577 & -7.904406 \\
2.6 & -7.884429 & -7.909037 & -7.911611 & -7.912994 \\
2.7 & -7.889869 & -7.919604 & -7.920041 & -7.922862 \\
2.8 & -7.895509 & -7.931308 & -7.928942 & -7.933881 \\
2.9 & -7.901382 & -7.943732 & -7.938379 & -7.945780 \\
3.0 & -7.907513 & -7.956643 & -7.948419 & -7.958342 \\ \toprule
\end{tabular}
\end{center}
\end{table}

\section{Total energies for Polyyne}

\begin{table}[H]
\begin{center}
\caption{Total energy per unit cell (in Hartree) for polyyne at different scaling factors, s.}
\begin{tabular}{K{1cm} K{2.0cm} K{2.0cm} K{3.5cm} K{3.0cm}K{3.0cm}}    \toprule
\textbf{s}	&   HF	&      CCSD	&   o-DMET(CCSD) &  o-DMET(FCI)  &  sc-DMET(FCI) \\
\hline
0.900 & -10.072662 & -10.165207 & -10.172702 & -10.174560 & -10.174975\\
0.925 & -10.119612 & -10.217977 & -10.226407 & -10.228321 & -10.228504\\
0.950 & -10.155428 & -10.259773 & -10.268610 & -10.271075 & -10.271003\\
0.975 & -10.181643 & -10.292138 & -10.301541 & -10.304365 & -10.304032\\
1.000 & -10.199582 & -10.316413 & -10.329789 & -10.329528 & -10.328920\\
1.025 & -10.210399 & -10.333763 & -10.344269 & -10.347728 & -10.346832\\
1.050 & -10.215103 & -10.345213 & -10.356046 & -10.359980 & -10.358782\\
1.075 & -10.214577 & -10.351662 & -10.362784 & -10.367174 & -10.365662\\
1.100 & -10.209595 & -10.353906 & -10.365340 & -10.370091 & -10.368253\\
1.125 & -10.200842 & -10.352655 & -10.367346 & -10.369417 & -10.367240\\
1.150 & -10.188919 & -10.348544 & -10.359862 & -10.365756 & -10.363221\\
1.175 & -10.174358 & -10.342155 & -10.353399 & -10.359638 & -10.356740\\
1.200 & -10.157626 & -10.334037 & -10.346292 & -10.351526 & -10.348253\\
1.225 & -10.139138 & -10.324741 & -10.336559 & -10.341825 & -10.338172\\
1.250 & -10.119257 & -10.314909 & -10.325782 & -10.330888 & -10.326847\\
\toprule
\end{tabular}
\end{center}
\end{table}

\begin{table}[H]
\begin{center}
\caption{CCSD total energy per unit cell (in Hartree) for 1D-H and polyyne using the $\Gamma$-centered and Z-centered \textbf{k}-mesh.}
\begin{tabular}{*3c}    \toprule
& \multicolumn{2}{c}{\textbf{k}-mesh}  \\
& $\Gamma$-centered & Z-centered\\\midrule
1D-H   & \multicolumn{1}{r}{-1.22084}  & \multicolumn{1}{r}{-1.22084} \\ 
Polyyne  & \multicolumn{1}{r}{-10.94416}  & \multicolumn{1}{r}{-10.94415} \\ \toprule
\end{tabular}
\end{center}
\end{table}